\documentclass[aps,prd,nofootinbib,twocolumn,superscriptaddress,preprintnumbers,balancelastpage,longbibliography]{revtex4-1}

\usepackage[utf8]{inputenc}
\usepackage{amsmath,amssymb,mathtools,bm}
\usepackage{graphicx, color, hepunits}
\usepackage[dvipsnames]{xcolor}
\usepackage{float}
\usepackage{multirow}
\usepackage{braket}
\usepackage{hyperref} 
\usepackage{caption}
\usepackage[lofdepth,lotdepth,justification=RaggedRight]{caption}

\usepackage{subcaption}

\hypersetup{
    colorlinks=true,      
    linkcolor=blue,        
    citecolor=blue,        
    filecolor=magenta,     
    urlcolor=blue          
}
\usepackage[utf8]{inputenc}
\usepackage[english]{babel}
\usepackage{lipsum}
\usepackage{mathrsfs}

\let\vec\mathbf
\usepackage{tensor}

\def\beq{\begin{equation}}
\def\eeq{\end{equation}}

\def\Beq{\begin{equation}\begin{aligned}}
\def\Eeq{\end{aligned}\end{equation}}

\def\bea{\begin{eqnarray}}
\def\eea{\end{eqnarray}}

\def\beq{\begin{equation}}
\def\eeq{\end{equation}}
\def\bea{\begin{eqnarray}}
\def\eea{\end{eqnarray}}

\DeclareMathOperator{\sech}{sech}

\newcommand{\PrOp}{\hat P_r}
\newcommand{\dr}{\frac{d}{dr}}

\begin{document}

\title{The Electroweak Sphaleron Revisited: \texorpdfstring{\\}{ }
I. Static Solutions, Energy Barrier, and Unstable Modes}

\author{Konstantin T. Matchev}
\affiliation{Department of Physics and Astronomy, University of Alabama, Tuscaloosa, AL 35487, USA}

\author{Sarunas Verner}
\affiliation{Institute for Fundamental Theory, Physics Department, University of Florida, Gainesville, FL 32611, USA}

\vspace{0.5cm}

\date{\today}

\begin{abstract}
The electroweak sphaleron is a static, unstable solution of the Standard Model classical field equations, representing the energy barrier between topologically distinct vacua. In this work, we present a comprehensive updated analysis of the sphaleron using current Standard Model parameters with the physical Higgs boson mass of $m_H = 125.1$ GeV and $m_W = 80.4$ GeV, rather than the $m_H = m_W$ approximation common in earlier studies. The study includes: (i) a complete derivation of the $SU(2)\times U(1)$ electroweak Lagrangian and field equations without gauge fixing constraints, (ii) high-precision numerical solutions for the static sphaleron configuration yielding a sphaleron energy $E_{\rm{sph}} \simeq 9.1$ TeV, (iii) an analysis of the minimum energy path in field space connecting the sphaleron to the vacuum (a 1D ``potential barrier" as a function of Chern–Simons number), and (iv) calculation of the sphaleron single unstable mode with negative eigenvalue $\omega^2_{-} = -2.7m^2_W$, providing analytical fits for its eigenfunction. We find that using the measured Higgs mass modifies the unstable mode frequency, with important implications for baryon number violation rates in both early universe cosmology and potential high-energy collider signatures. These results provide essential input for accurate lattice simulations of sphaleron transitions and precision calculations of baryon number violation processes.
\end{abstract}

\maketitle

%%%%%%%%%%%%%%%%%%%%%%%%%%%%%%%%%%%%%%%%%%%%%%%%%%%%%%%%%%%%%%%%%%%%%
\section{Introduction}
\label{sec:intro}
%%%%%%%%%%%%%%%%%%%%%%%%%%%%%%%%%%%%%%%%%%%%%%%%%%%%%%%%%%%%%%%%%%%%%
Nonperturbative electroweak processes can violate baryon ($B$) and lepton ($L$) number through the chiral anomaly \cite{Adler:1969gk, Bell:1969ts}. This was first demonstrated by 't Hooft \cite{tHooft:1976rip, tHooft:1976snw}, who showed that instanton solutions of the SU(2) gauge theory induce processes that change $B$ and $L$. At zero temperature such processes are exponentially suppressed by a factor of $\exp(-4\pi/\alpha_W) \sim 10^{-161}$, but Kuzmin, Rubakov, and Shaposhnikov pointed out that at high temperatures in the early Universe these baryon-number violating processes would occur at an unsuppressed rate \cite{Kuzmin:1985mm}. These transitions are instead governed by the so-called sphaleron: a static, unstable saddle-point configuration of the classical field equations that sits at the top of the energy barrier separating vacua of different Chern-Simons (CS) number. The electroweak sphaleron thus plays a pivotal role as the critical configuration mediating anomalous $B+L$ violation in the Standard Model (SM) at high temperatures.

The seminal work of Klinkhamer and Manton~\cite{Klinkhamer:1984di} identified this static saddle-point solution to the classical field equations of the Weinberg-Salam model in the limit where the weak mixing angle $\theta_W$ vanishes. This limiting case offers significant simplification: the $U(1)_{\rm Y}$ hypercharge gauge field completely decouples, reducing the theory to an $SU(2)_{\rm W}$ gauge theory coupled to a scalar Higgs doublet. Klinkhamer and Manton showed that the energy functional varies smoothly with $\theta_W$, such that finite but small values of the mixing angle contribute only perturbative corrections to the sphaleron configuration. The electroweak vacuum is topologically degenerate: configurations can be characterized by an integer Chern–Simons number $N_{\rm CS}$, and sphalerons correspond to configurations with half-integer values $N_{\rm CS}=n+1/2$ (where $n$ is an integer)~\cite{Manton:2004tk}. The sphaleron with $N_{\rm CS}=1/2$ is positioned at the peak of the energy barrier separating the vacua with $N_{\rm CS}=0$ and $N_{\rm CS}=1$.

More recently, it has been argued that sphaleron-like processes may also occur in high-energy collisions if sufficient energy is concentrated into the electroweak gauge and Higgs sectors. The possibility of sphaleron-mediated baryon and lepton number violation at high-energy colliders has been explored in~\cite{Gibbs:1994cw, Ringwald:2002sw, Bezrukov:2003er, Bezrukov:2003qm, Tye:2015tva}. In particular, Tye and Wong proposed a Bloch-wave picture in which the periodicity of the CS number leads to band structures in the transition probability, suggesting that above a certain energy threshold ($E \gtrsim E_{\rm sph}$), baryon number violation could proceed unsuppressed \cite{Tye:2015tva}. Subsequent work extended this framework to incorporate realistic parton distributions and collider observables, providing detailed cross-section estimates for sphaleron-induced transitions at the LHC and future colliders~\cite{Ellis:2016dgb, Ellis:2016ast}. These developments have renewed interest in understanding the precise structure, energy, and instability properties of the sphaleron, particularly in light of the precisely measured Higgs boson mass. 

Despite their theoretical significance, the quantitative characterization of the sphalerons has seen surprisingly few updates since the late 1980s. Most early works preceded the discovery of the Higgs boson and often assumed an approximate mass degeneracy $m_H = m_W$ for simplicity. Given that the Higgs mass is now known to be $m_H \simeq 125.1$ GeV (significantly larger than $m_W \simeq 80.4$ GeV) \cite{ParticleDataGroup:2024cfk}, it is important to revisit the sphaleron with physically accurate SM parameters. Furthermore, the CMS collaboration has recently performed dedicated searches for anomalous multi-fermion events consistent with sphaleron-induced $\Delta N_{\rm CS} = \pm 1$ transitions \cite{CMS:2018ozv}, making precision calculations of sphaleron properties directly relevant to current experimental efforts. If the Higgs were significantly heavier than its measured value, the sphaleron energy would increase and the static solution would deform, potentially altering the baryon number violation rate. However, the measured $125$ GeV Higgs ensures we are in the regime where the original sphaleron solution (symmetric under the full $SU(2)$ custodial symmetry) remains valid and dominant, though with quantitatively different properties than assumed in earlier works.

In this paper, we present an updated and comprehensive analysis of the electroweak sphaleron. We derive the full electroweak Lagrangian and classical field equations in their general form, without fixing a gauge or making the $SU(2)$ spherical ansatz \cite{Ratra:1987dp, Akiba:1988ay}. This provides a clear starting point and highlights all contributions (gauge, Higgs, and interactions) prior to fixing the gauge. Using a spherically symmetric $SU(2)$ ansatz (justified by the small mixing angle approximation and consistent with lattice studies), we obtain detailed numerical solutions for the sphaleron profile functions with current values of the Higgs and W-boson masses. From this, we compute the sphaleron energy $E_{\rm sph} \simeq 9.1$ TeV, in agreement with \cite{Tye:2015tva}.

We determine the static minimum-energy path in configuration space connecting the sphaleron to the vacuum. This one-dimensional ``potential barrier" allows us to interpret the sphaleron as the peak of a potential $V(N_{\rm CS})$, updating the classic barrier profiles of Akiba, Kikuchi, and Yanagida \cite{Akiba:1988ay}. We then perform a linear stability analysis around the sphaleron, confirming a single negative eigenvalue $\omega^2_{-} = -2.7m^2_W$ in the small-fluctuation spectrum. The corresponding eigenfunction is computed with analytical fits for its radial profile. Physically, this unstable mode corresponds to motion along the $N_{\rm CS}$ direction: perturbing the sphaleron in this mode causes it to ``decay" toward one of the vacua. The magnitude of the negative eigenvalue directly impacts the prefactor for the sphaleron transition rate in thermal environments, making this result particularly important for early Universe cosmology.

More crucially, sphaleron transitions in the early Universe play a fundamental role in baryogenesis~\cite{Cohen:1993nk, Rubakov:1996vz, Trodden:1998ym, Morrissey:2012db, Shaposhnikov:1987tw}. Because they violate $(B + L)$ but conserve $(B - L)$, sphalerons can convert a preexisting lepton asymmetry into a baryon asymmetry, as in standard leptogenesis scenarios \cite{Fukugita:1986hr, Davidson:2008bu}. Our results can help refine the condition for sphaleron freeze-out after the electroweak epoch. State-of-the-art lattice simulations (with $m_H = 125$ GeV) indicate that the electroweak phase transition is an analytic crossover at $T_c \simeq 159$ GeV~\cite{Manton:1983nd, Klinkhamer:1990fi, Kunz:1992uh, DOnofrio:2014rug}. Above this temperature, anomalous $B + L$ violating processes are rapid, while below $T_c$, the sphaleron rate drops steeply. Sphaleron processes are estimated to freeze out at a temperature $T_* \simeq 132$ GeV, below which baryon violation becomes negligible. This is crucial for low-scale leptogenesis scenarios, which generate a lepton asymmetry around the electroweak scale – the asymmetry can partially convert to baryon number, and then sphalerons turn off, preserving the baryon excess. Conversely, the SM crossover is not strong enough to prevent sphalerons from erasing any baryon asymmetry produced earlier than $T \simeq 130$ GeV, necessitating new physics for electroweak baryogenesis.

The structure of this paper is as follows: In Section~\ref{sec:fieldequations}, we derive the $SU(2)$ spherical symmetry ansatz and the resulting classical field equations. Section~\ref{sec:sphstaticminimum} presents our numerical approach for finding static sphaleron solutions and the boundary conditions employed. In Section~\ref{sec:staticminimsphandvac}, we analyze the minimum energy path between topologically distinct vacua. Section~\ref{sec:eigenvalueequations} focuses on determining the unstable mode and its properties. In Section~\ref{sec:1dbarrier}, we develop a one-dimensional effective model for sphaleron transitions. Finally, in Section~\ref{sec:conclusions}, we summarize our key findings and discuss their implications for both early Universe cosmology and high-energy collider physics.

%%%%%%%%%%%%%%%%%%%%%%%%%%%%%%%%%%%%%%%%%%%%%%%%%%%%%%%%%%%%%%%%%%%%%
\section{\boldmath \texorpdfstring{$SU(2)$}{SU(2)} Spherical Symmetry and Classical Field Equations}
\label{sec:fieldequations}
%%%%%%%%%%%%%%%%%%%%%%%%%%%%%%%%%%%%%%%%%%%%%%%%%%%%%%%%%%%%%%%%%%%%%
We begin our analysis by considering the $SU(2)_W \times U(1)_Y$ electroweak theory, setting the foundation for our investigation of the sphaleron solution.
Later, we will work in the limit where the weak mixing angle $\theta_W \rightarrow 0$. In this approximation, the $U(1)_Y$ hypercharge gauge field decouples, allowing us to focus on the $SU(2)_W$ weak interaction gauge fields $W^a_\mu(x)$ coupled to a complex scalar doublet $\Phi(x)$ and left-handed fermion doublets $\Psi^{(f)}_L = (q^{f}_L, l^f_L)$, with $f = 1,2,3$ labeling fermion families.

Multiple studies have confirmed the validity of this approximation~\cite{Manton:1983nd, Klinkhamer:1990fi, Kunz:1992uh}, showing that including the $U(1)$ hypercharge coupling with the physical value $\sin^2\theta_W \simeq 0.23$ reduces the sphaleron energy by approximately 1\%. We therefore employ the simplified estimate $E_{\rm{sph}} \simeq 9.0~\rm{TeV}$, which provides sufficient accuracy for both theoretical explorations and phenomenological analyses of processes that violate baryon number conservation.

To formulate our analysis precisely, we now present the full mathematical framework underlying the sphaleron configuration in the electroweak theory. We begin by considering the following bosonic fields in the $SU(2)_{W} \times U(1)_Y$ electroweak theory:
\begin{equation}
    \label{eq:bosonlagr}
    \mathcal{L} \; = \;  -\frac{1}{4} \left(W_{\mu \nu}^a \right)^2 - \frac{1}{4} B_{\mu \nu}^2 + (D_{\mu} \Phi)^{\dagger} (D^{\mu} \Phi) - V(\Phi)  \, ,
\end{equation}
where $W_{\mu}^a$ are the $SU(2)$ gauge bosons with field strengths $W_{\mu \nu}^{a} = \partial_{\mu} W_{\nu}^a - \partial_{\nu}W_{\mu}^a + g \varepsilon^{abc} W_{\mu}^b W_{\nu}^c$, where $\varepsilon^{abc}$ is the totally antisymmetric structure constant, and $B_{\mu}$ is the hypercharge gauge boson with $B_{\mu \nu} = \partial_{\mu} B_{\nu} - \partial_{\nu} B_{\mu}$. The covariant derivative acting on the Higgs doublet is given by
\begin{equation}
    D_{\mu} \Phi\; = \; \partial_{\mu}\Phi - \frac{1}{2}ig W_{\mu}^a \sigma^a \Phi - \frac{1}{2}i g' B_{\mu} \Phi\, ,
\end{equation}
where $g$ is the $SU(2)_{W}$ coupling, $g'$ is the $U(1)_{Y}$ coupling, and $\sigma^a$ are the Pauli matrices. The factor of $1/2$ in the $B_{\mu}$ coupling comes from the Higgs doublet hypercharge $Y = 1/2$. The Higgs potential takes the standard form:
\begin{equation}
    V(\Phi) \; = \; -\mu^2 |\Phi|^2 + \lambda |\Phi|^4 \, ,
\end{equation}
with mass parameter $\mu$ and self-coupling $\lambda$. After spontaneous symmetry breaking, the Higgs field acquires a vacuum expectation value $v = \mu/\sqrt{\lambda} \simeq 246~\rm{GeV}$, giving rise to the $W$-boson mass $m_W = gv/2$ and the Higgs boson mass $m_H = \sqrt{2\lambda}v$. The electroweak mixing angle is defined as $\tan{\theta_{W}} = g'/g$.

Using the Euler-Lagrange equations, we derive the full field equations from the Lagrangian \eqref{eq:bosonlagr}:
\begin{align}
    D_{\mu} D^{\mu} \Phi &\; = \; -\frac{\partial V(\Phi)}{\partial \Phi^{\dagger}} \, , \\
    D^{\nu} W_{\mu \nu}^a &\; = \; -i\frac{g}{2}\left[\Phi^{\dagger} \sigma^a \left(D_{\mu} \Phi \right) - \left(D_{\mu} \Phi \right)^{\dagger} \sigma^a \Phi \right] \, , \\
     \partial^{\nu} B_{\mu \nu} &\; = \; -i\frac{g'}{2} \left[\Phi^{\dagger} \left(D_{\mu} \Phi \right) - \left(D_{\mu} \Phi \right)^{\dagger} \Phi \right] \, ,
\end{align}    
where we define the covariant derivative of the field strength as:
\begin{equation}
\begin{aligned}
     D^{\nu} W_{\mu \nu}^a &\; = \; \partial^{\nu} W_{\mu \nu}^a + g \varepsilon^{abc}W_b^{\nu} W_{\mu \nu c} \, . 
\end{aligned}    
\end{equation}

As we set $g' = 0$ to analyze the sphaleron in the $\theta_W \rightarrow 0$ limit, the $U(1)_Y$ field decouples from the theory. After spontaneous symmetry breaking, the Higgs potential can be rewritten as:
\begin{equation}
V(|\Phi|) = \lambda\left(|\Phi|^2 - \frac{v^2}{2}\right)^2,
\label{eq:higgsvev}
\end{equation}
which has a minimum at $|\Phi| = v/\sqrt{2}$. The field equations simplify to:
\begin{align}
    \label{eq:eom1}
    D_{\mu} D^{\mu} \Phi &\; = \; -2 \lambda \left(|\Phi|^2 - \frac{v^2}{2} \right) \Phi \, , \\
    \label{eq:eom2}
    D^{\nu} W_{\mu \nu}^a &\; = \; -i \frac{g}{2} \left[\Phi^{\dagger} \sigma^a \left(D_{\mu} \Phi \right) - \left(D_{\mu} \Phi \right)^{\dagger} \sigma^a \Phi \right] \, .  
\end{align}    

To capture the essential features of the sphaleron configuration, we introduce a general spherically symmetric ansatz that respects the underlying symmetries of the system while providing a tractable mathematical framework ~\cite{Ratra:1987dp, Akiba:1988ay}:
\begin{align}
\label{eq:sphans1}
W_0^a(x) \; &=\; \frac{1}{g}\,G(r,t)\,\frac{x_a}{r}\,,
\\
\label{eq:sphans2}
    W_j^a(x)  &\; = \;\frac{1}{g} \left[\frac{(f_A(r,t) - 1)}{r^2} \varepsilon_{jam} x_m \right.~\nonumber \\ 
    &\left.  + \frac{f_B(r,t)}{r^3}\left(r^2 \delta_{ja} - x_j x_a \right) + \frac{f_C(r,t)}{r^2}x_jx_a \right] \, ,
\\
\label{eq:sphans3}
\Phi(x) \; &=\; \frac{v}{\sqrt{2}}\!
\left[\,H(r,t)
   + i\,K(r,t)\,\frac{\boldsymbol{\sigma}\!\cdot\!\mathbf{x}}{r}
\right]
\begin{pmatrix}
 0 \\ 1
\end{pmatrix} \, .
\end{align}
where $r = |\mathbf{x}|$ is the radial coordinate and $\boldsymbol{\sigma} \cdot \mathbf{x} = \sigma^a x_a$. This ansatz is particularly suitable for describing the sphaleron as it incorporates the hedgehog configuration characteristic of topologically non-trivial gauge field solutions.
    
Substituting these expressions into the field equations \eqref{eq:eom1}-\eqref{eq:eom2}, we obtain a system of six coupled partial differential equations for the functions $G(r,t)$, $f_A(r,t)$, $f_B(r,t)$, $f_C(r,t)$, $H(r,t)$, and $K(r,t)$. These equations fully characterize the dynamics of the spherically symmetric configurations in our model:
\begin{widetext}
\begin{equation}
\begin{aligned}
\label{eq:fa1}
\ddot{f}_A - f_A'' 
&+ \frac{1}{r^2} \left(f_A^2 + f_B^2 - 1 \right) f_A 
+ m_W^2 \left[(H^2+ K^2)f_A + K^2 - H^2 \right] \\
&+ f_A f_C^2 - 2 f_B' f_C - f_B f_C' - f_A G^2 +  2 G \dot{f}_B + f_B \dot{G} = 0 \, .
\end{aligned}
\end{equation}
\begin{equation}
\begin{aligned}
\label{eq:fb1}
\ddot{f}_B - f_B'' 
&+ \frac{1}{r^2} \left(f_A^2 + f_B^2 - 1 \right) f_B 
+ m_W^2 \left[(H^2+ K^2)f_B - 2HK \right] \\
&+ f_B f_C^2  + 2 f_A' f_C + f_A f_C'  - f_B G^2 - 2 G \dot{f}_A  - f_A \dot{G} 
= 0 \, .
\end{aligned}
\end{equation}
\begin{equation}
\begin{aligned}
\label{eq:fc1}
\ddot{f}_C 
+ \frac{2}{r^2} \left(f_A^2 + f_B^2 \right) f_C 
&+ m_W^2 \left(H^2 + K^2 \right) f_C 
+ 2m_W^2 \left(H'K - HK' \right) 
+ \frac{2}{r^2} \left(f_A' f_B - f_A f_B' \right) \\
& + \frac{1}{r}\left(f_B G^2 + 2 G \dot{f}_A +f_A \dot{G} \right) - \dot{G}'  = 0 \, .
\end{aligned}
\end{equation}
\begin{align}
\label{eq:h1}
    &\ddot{H} - \frac{1}{r}(r H)'' +\frac{1}{2r^2} \left(f_A^2 + f_B^2 + 1 \right)H - \frac{1}{r^2} \left(H f_A + K f_B \right) + \frac{1}{2}m_H^2 \left(H^2 + K^2 - 1 \right) H ~\nonumber \\
    &- \frac{1}{r} K f_C + \frac{1}{4} H (f_C^2 - G^2) - K'f_C + \dot{K} G - \frac{1}{2}K (f_C' - \dot{G}) \; = \; 0 \, ,
\end{align}
\begin{align}
\label{eq:k1}
    &\ddot{K} - \frac{1}{r}(r K)'' +\frac{1}{2r^2} \left(f_A^2 + f_B^2 + 1 \right)K + \frac{1}{r^2} \left(K f_A - H f_B \right) + \frac{1}{2}m_H^2 \left(H^2 + K^2 - 1 \right) K~\nonumber \\
    &+\frac{1}{r} H f_C + \frac{1}{4} K (f_C^2 - G^2) + H' f_C  - \dot{H}G + \frac{1}{2}H (f_C' - \dot{G})\; = \; 0 \, ,
\end{align}
\begin{equation}
\begin{aligned}
\label{eq:g1}
&\frac{1}{r}(r G)''  - \dot{f}_C' - \frac{2}{r} \dot{f}_C 
- \frac{2}{r^2}(f_A^2 + f_B^2) G 
+ \frac{2}{r^2} (f_A \dot{f}_B -  f_B \dot{f}_A ) + 2 m_W^2 ( H \dot{K} -  K \dot{H} )  - m_W^2 (H^2 + K^2) G
 \; = \; 0 \, .
\end{aligned}
\end{equation}
\end{widetext}

An important feature of the spherically symmetric ansatz is that it preserves a residual U(1) gauge symmetry, which is a subgroup of the original SU(2). As shown by Ref.~\cite{Ratra:1987dp}, this symmetry acts nontrivially on the fields as:
\begin{align}
\begin{pmatrix}
G \\
f_C
\end{pmatrix}
&\rightarrow
\begin{pmatrix}
G \\
f_C
\end{pmatrix}
+
\begin{pmatrix}
\dot{\theta}(r,t) \\
\theta'(r,t)
\end{pmatrix} \, ,  \\
\begin{pmatrix}
f_A \\
f_B
\end{pmatrix}
&\rightarrow
\begin{pmatrix}
\cos\theta(r,t) & -\sin\theta(r,t) \\
\sin\theta(r,t) & \cos\theta(r,t)
\end{pmatrix}
\begin{pmatrix}
f_A \\
f_B
\end{pmatrix} \, , \\
\begin{pmatrix}
H \\
K
\end{pmatrix}
&\rightarrow
\begin{pmatrix}
\cos\left(\frac{\theta(r,t)}{2}\right) & -\sin\left(\frac{\theta(r,t)}{2}\right) \\
\sin\left(\frac{\theta(r,t)}{2}\right) & \cos\left(\frac{\theta(r,t)}{2}\right)
\end{pmatrix}
\begin{pmatrix}
H \\
K
\end{pmatrix} \, . 
\end{align}
Here, the prime denotes a radial derivative, i.e., $\theta'(r,t) \equiv \partial \theta(r,t)/\partial r$, and the dot denotes a time derivative, i.e., $\dot{\theta}(r,t) \equiv \partial \theta(r,t)/\partial t$. The function $\theta(r,t)$ parametrizes local $\mathrm{SU}(2)$ gauge transformations that preserve the structure of the ansatz. These transformations define a residual $\mathrm{U}(1)$ gauge freedom within the space of spherically symmetric configurations. Importantly, no explicit gauge fixing is required at this stage: the residual gauge symmetry is identified by analyzing how the ansatz transforms under infinitesimal gauge rotations. At a later stage, one may choose a specific representative within the $\mathrm{U}(1)$ gauge orbit, for example, by taking a time-independent transformation with $\bar{\theta}(r) = -\pi$, to obtain a localized configuration with nontrivial topological charge, corresponding to the sphaleron. This choice is not necessary for defining the ansatz, but it selects a convenient gauge representative within the class of physically equivalent configurations.

The total energy functional, which can be derived from the Lagrangian~(\ref{eq:bosonlagr}), is given by:
\begin{equation}
    \label{eq:totalenergy}
    E \; = \; E_{\rm kin} + E_{\rm static} \, , 
\end{equation}
where the kinetic energy contribution is given by
\begin{equation}
\begin{aligned}
    E_{\rm kin} & = \frac{4\pi}{g^2} \int_0^{\infty} dr \Biggl[ \dot{f}_A^2 \left( 1 + \frac{2f_B G}{\dot{f}_A}\right) \\
    &+ \dot{f}_B^2 \left( 1 - \frac{2f_A G}{\dot{f}_B}\right) + \frac{r^2}{2} \dot{f}_C^2 \left(1-2 \frac{G'}{\dot{f}_C} \right) \\
    &+ 2 m_W^2 r^2 \left( \dot{H}^2 + \dot{K}^2 + G \left(\dot{H}K - H\dot{K}\right)\right) \Biggr] \, ,
 \label{eq:kinen}   
\end{aligned}
\end{equation}
and the static energy contribution takes the form
\begin{equation}
\begin{aligned}
    & E_{\rm static} = \frac{4\pi}{g^2} \int_0^{\infty} dr \Biggl[ \left(f_A' + f_C f_B \right)^2 + \left(f_B' - f_C f_A \right)^2 \\
    &+ \frac{\left(f_A^2 + f_B^2 - 1 \right)^2}{2r^2} -\left(f_A^2 + f_B^2\right)G^2 - \frac{1}{2}r^2G'^2 \\
    & + 2 m_W^2 r^2 \bigg\{ \left(H' + \frac{1}{2}f_C K \right)^2 + \left(K' - \frac{1}{2} f_C H \right)^2 \\
    & + \frac{1}{2r^2} \left(H f_A + K f_B - H \right)^2 + \frac{1}{2r^2} \left(K f_A - H f_B + K \right)^2 \\
    & -\frac{1}{4} G^2(H^2 + K^2) \bigg\} \\
    & + \frac{(m_W m_H)^2}{2} r^2 \left(H^2 + K^2 -1 \right)^2 
    \Biggr] \, .
\label{eq:staten}   
\end{aligned}
\end{equation}
We note that these expressions are completely general and the gauge has not been fixed yet.

The total energy functional presented encapsulates all the essential dynamics of the field configuration without imposing any gauge-fixing conditions. Several important features can be observed in these expressions: 1) The prefactor $\frac{4\pi}{g^2} \equiv \frac{1}{\alpha_{\rm W}}$ sets the characteristic energy scale of the sphaleron, directly related to the electroweak scale and inversely proportional to the weak fine-structure constant. 2) Cross-terms coupling the gauge fields ($f_A$, $f_B$, $f_C$) with the Higgs components ($H$, $K$) reflect the essential non-Abelian character of the energy functional, highlighting how gauge and Higgs sectors interact non-trivially.
3) The kinetic energy includes terms involving the time evolution of all dynamical variables, crucial for understanding transitions between the sphaleron and vacuum configurations.

Therefore, our analysis addresses three fundamental aspects of the electroweak sphaleron:
\begin{enumerate}
    \item \textbf{Static Sphaleron Solution}: First, we will determine the static sphaleron configuration by minimizing $E_{\rm static}$ subject to appropriate boundary conditions. This saddle-point solution represents the minimum-energy barrier configuration between topologically distinct vacua.
    
    \item \textbf{Minimum Energy Path}: Next, we will compute the minimum energy path connecting the sphaleron to the vacuum configuration. This path, often called the ``most probable escape path," characterizes the optimal trajectory for baryon-number-violating processes and is crucial for determining transition rates in thermal environments.
    
    \item \textbf{Unstable Mode Analysis}: Finally, we identify and analyze the unstable mode of the sphaleron, which drives the system away from the saddle point toward either vacuum state. The eigenvalue associated with this mode determines the prefactor in the thermal transition rate, while its eigenvector characterizes how the sphaleron decays.
\end{enumerate}

%%%%%%%%%%%%%%%%%%%%%%%%%%%%%%%%%%%%%%%%%%%%%%%%%%%%%%%%%%%%%%%%%%%%%
\section{Sphaleron Solution and Static Minimum-Energy Path}
\label{sec:sphstaticminimum}
%%%%%%%%%%%%%%%%%%%%%%%%%%%%%%%%%%%%%%%%%%%%%%%%%%%%%%%%%%%%%%%%%%%%%

%%%%%%%%%%%%%%%%%%%%%%%%%%%%%%%%%%%%%%%%%%%%%%%%%%%%%%%%%%%%%%%%%%%%%
\subsection{Static Sphaleron Equations of Motion}
%%%%%%%%%%%%%%%%%%%%%%%%%%%%%%%%%%%%%%%%%%%%%%%%%%%%%%%%%%%%%%%%%%%%%
In this section, we derive and analyze the static sphaleron solution that represents the energy barrier between topologically distinct vacua. We begin by simplifying the general field equations for static configurations, establish appropriate boundary conditions, and develop a numerical framework for obtaining precise solutions with current Standard Model parameters.

For static configurations, all time derivatives vanish and the time component of the gauge field must be zero, i.e., $G(r,t) = 0$. Additionally, we adopt the radial gauge condition $x^i W^a_i = 0$, which corresponds to setting $f_C(r,t) = 0$. Under these conditions, the equations of motion~(\ref{eq:fa1})-(\ref{eq:g1}) reduce to the following system of coupled ordinary differential equations~\cite{Akiba:1988ay}:
\begin{equation}
\begin{aligned}
\label{eq:fa2}
f_A'' &- \frac{1}{r^2} \left(f_A^2 + f_B^2 - 1 \right) f_A \\
&- m_W^2 \left[(H^2+ K^2)f_A + K^2 - H^2 \right]  \; = \; 0 \, ,
\end{aligned}
\end{equation}
\begin{equation}
\begin{aligned}
    \label{eq:fb2}
   f_B'' &- \frac{1}{r^2} \left(f_A^2 + f_B^2 - 1 \right) f_B \\
   &- m_W^2 \left[(H^2+ K^2)f_B -2HK\right] \; = \; 0  \, ,
\end{aligned}
\end{equation}
\begin{equation}
    \label{eq:const2}
     f_A' f_B - f_A f_B' + m_W^2 r^2 \left(H'K - HK'\right)\; = \; 0  \, ,
\end{equation}
\begin{equation}
\begin{aligned}
    \label{eq:h2}
   \frac{1}{r}(r H)''  &-\frac{1}{2r^2} \left(f_A^2 + f_B^2 + 1 \right)H + \frac{1}{r^2} \left(H f_A + K f_B \right)\\
    & - \frac{1}{2}m_H^2 \left(H^2 + K^2 - 1 \right) H \; = \; 0 \, ,
\end{aligned}
\end{equation}
\begin{equation}
\begin{aligned}
    \label{eq:k2}
     \frac{1}{r}(r K)'' &- \frac{1}{2r^2} \left(f_A^2 + f_B^2 + 1 \right)K - \frac{1}{r^2} \left(K f_A - H f_B \right) \\
     &- \frac{1}{2}m_H^2 \left(H^2 + K^2 - 1 \right) K  \; = \; 0 \, .
\end{aligned}
\end{equation}

These equations describe a static, spherically symmetric field configuration. The static energy functional for these field configurations simplifies to:
\begin{equation}
\begin{aligned}
    &E_{\rm sph} \; = \;  \frac{4\pi}{g^2} \int_0^{\infty} dr \Biggl[ \left(f_A'^2 + f_B'^2 + \frac{\left(f_A^2 + f_B^2 - 1 \right)^2}{2r^2} \right) \\
    & + 2 m_W^2 r^2 \bigg( H'^2 + K'^2 + \frac{1}{2r^2} \left(H f_A + K f_B - H \right)^2 \\
    & + \frac{1}{2r^2} \left(K f_A - H f_B + K \right)^2 \bigg) \\
    &+ \frac{(m_W m_H)^2}{2} r^2 \left(H^2 + K^2 -1 \right)^2 
    \Biggr] \, .
\end{aligned}
\end{equation}
We note that not all equations are independent. Equation~(\ref{eq:const2}) represents a constraint whose left-hand side corresponds to an integration constant for the other equations, which vanishes as $r \to \infty$. 
%%%%%%%%%%%%%%%%%%%%%%%%%%%%%%%%%%%%%%%%%%%%%%%%%%%%%%%%%%%%%%%%%%%%%
\subsection{Pure Gauge Configurations and Boundary Conditions}
\begin{figure*}[ht!]
    \centering
    
    % --- Top row ---
    \begin{subfigure}[b]{0.49\textwidth}
        \centering
        \includegraphics[width=\textwidth]{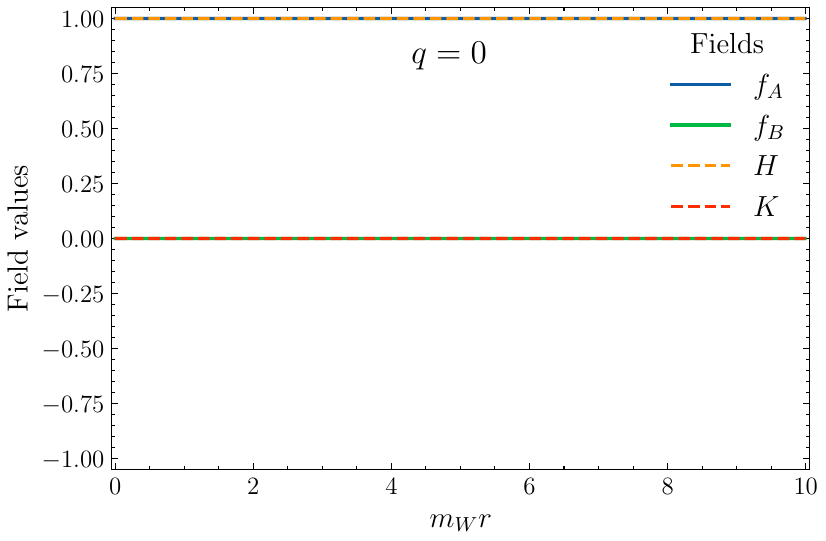}
    \end{subfigure}
    \hfill
    \begin{subfigure}[b]{0.49\textwidth}
        \centering
        \includegraphics[width=\textwidth]{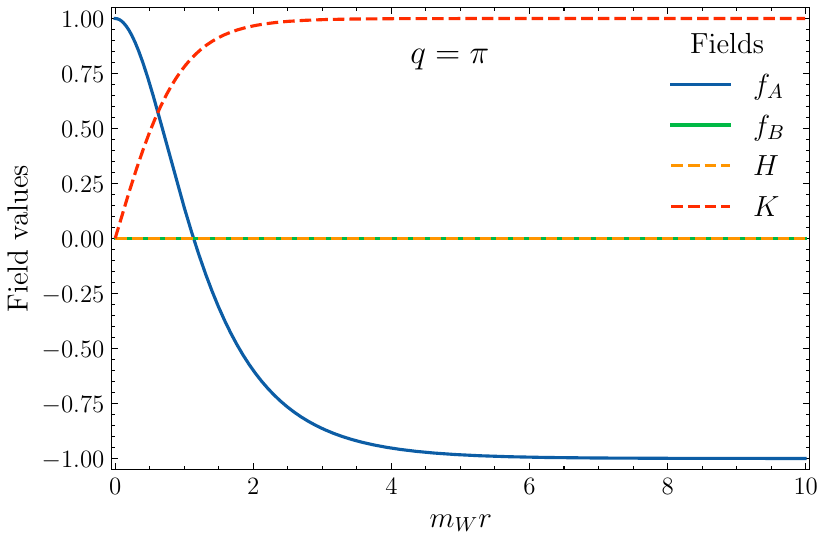}
    \end{subfigure}
    
    \vspace{0.5cm}
    
    % --- Bottom row ---
    \begin{subfigure}[b]{0.49\textwidth}
        \centering
        \includegraphics[width=\textwidth]{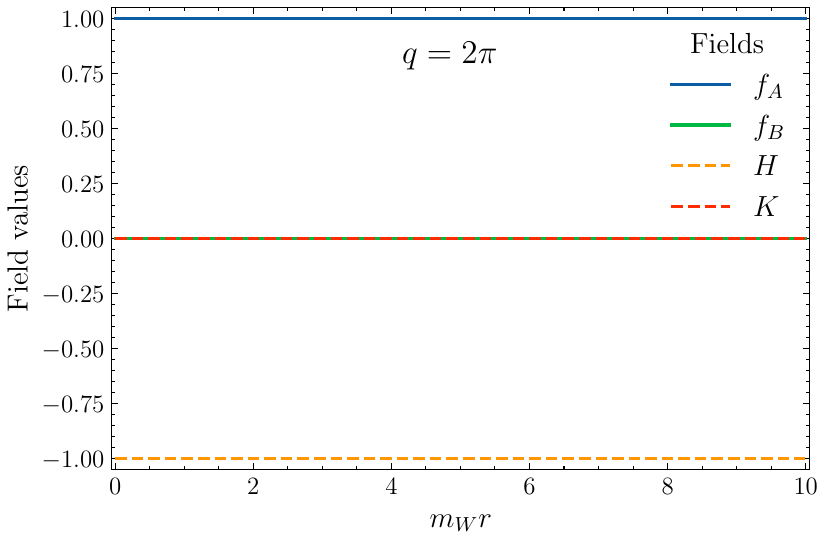}
    \end{subfigure}
    \hfill
    \begin{subfigure}[b]{0.49\textwidth}
        \centering
        \includegraphics[width=\textwidth]{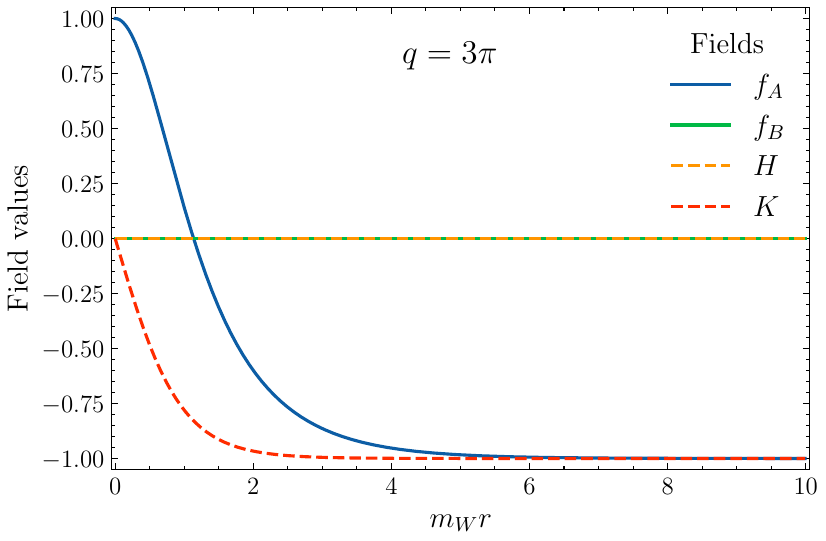}
    \end{subfigure}
    
    \caption{Field profiles of the sphaleron configuration and vacuum solutions for different values of the parameter $q$, which interpolates between topologically distinct vacua in $SU(2)_{\rm W}$ gauge theory. Each panel shows the radial dependence of the gauge field functions $f_A$, $f_B$, and the scalar fields $H$, $K$ as functions of $m_W r$. The configurations at $q = 0 $, $2\pi$ (left panels) correspond to vacuum solutions, while $q = \pi$, $ 3\pi$ (right panels) correspond to sphaleron configurations centered at the top of the energy barrier separating neighboring vacua.}
    \label{fig:sphaleronandvacua}
\end{figure*}

%%%%%%%%%%%%%%%%%%%%%%%%%%%%%%%%%%%%%%%%%%%%%%%%%%%%%%%%%%%%%%%%%%%%%
Following the discussion in Ref.~\cite{Akiba:1988ay}, we derive general expressions for sphaleron and vacuum boundary conditions incorporating pure gauge configurations. In the radial gauge, the gauge potential has no radial component. This property implies that at large radius $r \to \infty$, a finite-energy field configuration must asymptotically approach a pure gauge. Therefore, we can solve the $SU(2)_{\rm W}$ equations of motion as $r \rightarrow \infty$ by using:
\begin{align}
\sigma^a W_j^a(\infty) \; = \; -\frac{2i}{g} \, \partial_j U U^{-1} \, , 
 \quad \Phi(\infty) \; = \; \frac{v}{\sqrt{2}} \, U
\begin{pmatrix}
0 \\
1
\end{pmatrix} \, , 
\end{align}
where the $SU(2)_{\rm W}$ group element $U(r)$ is given by:
\begin{equation}
U(r) \; = \; \exp\left( \frac{i}{2} q \, \frac{\boldsymbol{\sigma} \cdot \mathbf{r}}{r} \right) \, .
\end{equation}
Here $q$ is a free parameter that we vary between $0$ and $4\pi$.\footnote{We vary $q$ from $0$ to $4\pi$ rather than from $0$ to $2\pi$ because the sphaleron solution at $q = 3\pi$ differs from the one at $q = \pi$ by a sign change in the $K$ field component. This behavior is illustrated in Fig.~\ref{fig:sphaleronandvacua}.} The pure gauge condition also implies that $W_{ij}^a (r\to\infty) = 0$. Physically, this can be understood as a consequence of the finite energy constraint, which forces the fields to approach a ``pure gauge + vacuum expectation value (VEV)" configuration at spatial infinity. The equations of motion are then automatically satisfied in this asymptotic region, as the fields have already reached a minimum energy configuration. More generally, the parameter $q$ in the group element $U$ determines which topological vacuum is approached at infinity, leading to different possible Chern-Simons or winding numbers. We use different values of $q$ to identify and characterize vacuum and sphaleron solutions.

We now proceed to define the boundary conditions as a function of $q$. At $r = 0$, the regularity conditions from the equations of motion~(\ref{eq:fa2})--(\ref{eq:k2}) give:
\begin{align}
f_A &\simeq 1 + \gamma_{A} (m_W r)^2 \, , \\
f_B &\simeq -\frac{1}{3} \gamma_H \gamma_K (m_W r)^3 \, , \\
H &\simeq \gamma_H + \frac{1}{12} \gamma_H (\gamma_H^2 - 1)(m_H r)^2 \, , \\
K &\simeq \gamma_K (m_W r) \, ,
\end{align}
and the asymptotic behavior at large $r$ is:
\begin{align}
f_A &\simeq \cos q + (\delta_A \cos q - \delta_B \sin q) \exp(-m_W r) \, , \\
f_B &\simeq \sin q + (\delta_A \sin q + \delta_B \cos q) \exp(-m_W r) \, , \\
H &\simeq \cos\left(\frac{q}{2}\right) \left[1 + \frac{\delta_H}{m_W r} \exp(-m_H r)\right] \\
&+ \sin\left(\frac{q}{2}\right) \frac{\delta_B}{m_W^2 r^2} \exp(-m_W r) \, , \\
K &\simeq \sin\left(\frac{q}{2}\right) \left[1 + \frac{\delta_H}{m_W r} \exp(-m_H r)\right] \\
&- \cos\left(\frac{q}{2}\right) \frac{\delta_B}{m_W^2 r^2} \exp(-m_W r) \, ,
\end{align}
in agreement with Ref.~\cite{Akiba:1988ay}.

Using numerical fits to our solutions, we find that for the sphaleron configuration at $q = \pi$:
\begin{align}
    \gamma_A &= -1.16  \left(\frac{m_H}{m_W}\right)^{0.34} \simeq -0.998 \, , \\
    \gamma_K &= 0.87 \left(\frac{m_H}{m_W}\right)^{0.45} \simeq 1.062  \, , \\
    \delta_A &= -2.92 \left(\frac{m_W}{m_H}\right)^{-0.34} \simeq -2.512 \, ,
\end{align}
while the remaining constants ($\gamma_H$, $\delta_B$, $\delta_H$) are zero. In these expressions, we used the current SM parameters, with $m_H = 125.11 \, \rm{GeV}$ and $m_W = 80.36 \, \rm{GeV}$, leading to the mass ratio $m_H/m_W \simeq 1.556$.

We note that when scanning for solutions with $q = 3\pi$, the parameter $\gamma_K$ acquires a minus sign compared to the $q = \pi$ case. The sphaleron and vacuum solutions are illustrated in Fig.~\ref{fig:sphaleronandvacua}. We searched for solutions in the region $q \in [0, 4\pi]$ and found only sphaleron and vacuum solutions, consistent with the findings in Ref.~\cite{Akiba:1988ay}.
We find the sphaleron energy to be 
\begin{equation}
    E_{\rm{sph}} \simeq 9.08 \, \rm{TeV} \, ,
\end{equation}
in excellent agreement with the detailed numerical determination of Ref.~\cite{Tye:2015tva}, which reports $E_{\rm{sph}} \simeq 9.11 \, \rm{TeV}$.

%%%%%%%%%%%%%%%%%%%%%%%%%%%%%%%%%%%%%%%%%%%%%%%%%%%%%%%%%%%%%%%%%%%%%
\section{Static Minimum-Energy Path Between Sphaleron and Vacuum}
\label{sec:staticminimsphandvac}
%%%%%%%%%%%%%%%%%%%%%%%%%%%%%%%%%%%%%%%%%%%%%%%%%%%%%%%%%%%%%%%%%%%%%

%%%%%%%%%%%%%%%%%%%%%%%%%%%%%%%%%%%%%%%%%%%%%%%%%%%%%%%%%%%%%%%%%%%%%
\subsection{Topological Structure and Anomaly Equation}
%%%%%%%%%%%%%%%%%%%%%%%%%%%%%%%%%%%%%%%%%%%%%%%%%%%%%%%%%%%%%%%%%%%%%
Having established the static sphaleron solution in the previous section, we now focus on constructing and analyzing the minimum-energy path connecting the sphaleron to topologically distinct vacuum states. This path is crucial for understanding baryon number violation processes mediated by sphalerons in both early universe cosmology and high-energy particle collisions.

In the Standard Model (or any non-Abelian gauge theory with chiral fermions), the baryon number current $J_B^\mu$ is not conserved at the quantum level due to the chiral anomaly~\cite{Adler:1969gk, Bell:1969ts, tHooft:1976rip}. Instead, it satisfies an anomaly equation of the form
\begin{equation}
\partial_\mu J_B^\mu
\;=\;
\frac{N_f g^2}{32\pi^2} \operatorname{Tr}(W_{\mu\nu}\,\widetilde{W}^{\mu\nu}) \, ,
\label{eq:anomaly}
\end{equation}
where $N_f = 3$ is the number of fermion families, $\widetilde{W}^{a \mu\nu} = \tfrac12\,\varepsilon^{\mu\nu\rho\sigma}\,W_{\rho\sigma}^a$ is the field dual, and the trace is over the $SU(2)$ gauge indices. This equation connects fermion number violation to the topology of gauge field configurations. The right-hand side of Eq.~(\ref{eq:anomaly}) can be rewritten as the four-divergence of a current $K^\mu$, known as the Chern-Simons current:
\begin{equation}
\frac{g^2}{32\pi^2} \operatorname{Tr}(W_{\mu\nu}\,\widetilde{W}^{\mu\nu}) \; = \; \partial_\mu K^\mu \, ,
\label{eq:cscurrent}
\end{equation}
where $K^\mu$ is given by:
\begin{equation}
K^\mu \; = \; \frac{g^2}{16\pi^2}\varepsilon^{\mu\nu\rho\sigma}
\Bigl(
W^a_{\nu} \partial_{\rho} W^a_{\sigma}
-\tfrac{g}{3} \varepsilon^{abc} W_\nu^a W_\rho^b W_\sigma^c
\Bigr) \, .
\label{eq:kdivergence}
\end{equation}

When instanton solutions exist in Euclidean spacetime~\cite{Belavin:1975fg}, the integral of the anomaly over all spacetime gives a topological winding number:
\begin{equation}
    N \; = \; \frac{g^2}{32\pi^2} \int d^4x  \operatorname{Tr}(W_{\mu\nu}\,\widetilde{W}^{\mu\nu})  \, ,
    \label{eq:Nnumber}
\end{equation}
where $N$ takes only integer values. An instanton with winding number $N$ mediates a tunneling process between states with different baryon numbers: $\ket{n} \rightarrow \ket{n + N}$, where the change in baryon number is $\Delta B = 3N$ for the Standard Model with three fermion generations. Using Stokes' theorem, we can express the topological index in terms of the time component of the Chern-Simons current:
\begin{equation}
N(t_0) = \int d^3x\, K^0 \Big|_{t = t_0}^{t = -\infty} + \int_{-\infty}^{t_0} dt \int_S \vec{K} \cdot d\vec{S} \, ,
\end{equation}
where $S$ is the surface at spatial infinity. If $\vec{K}$ decreases sufficiently rapidly as $r \to \infty$, the surface integral vanishes, and we can define the Chern-Simons number at time $t_0$ as:
\begin{equation}
N_{\text{CS}}(t_0) = \int d^3x\, K^0(t_0) \, ,
\label{eq:NCS}
\end{equation}
assuming $K^0 = 0$ at $t = -\infty$ for simplicity.

In the radial gauge we have adopted, the topological number $N$ receives an additional contribution from a non-vanishing surface term~\cite{Akiba:1988ay,Klinkhamer:1984di}, leading to:
\begin{equation}
\begin{aligned}
& N = \int d^3x\, K^0 + \frac{q - \sin q}{2\pi} \\
&= \frac{1}{2\pi} \left( \int_{r=0}^{r = \infty} dr (f_A' f_B - f_A f_B') + f_B\Big|_{r = 0}^{r = \infty} \right) + \frac{q - \sin q}{2\pi} \, ,
\end{aligned}
\label{eq:Nradial}
\end{equation}
where $q$ is the parameter in the pure gauge configuration discussed in Section~\ref{sec:sphstaticminimum}. Importantly, $N$ is gauge-invariant and directly related to the baryon-number violation $\Delta B = 3N$ corresponding to three fermion families.

A vacuum (or ground state) of the $SU(2)_W$ theory has the Higgs field at its VEV magnitude, $|\Phi| = v/\sqrt{2}$, and the gauge field in a pure gauge configuration. Such configurations are characterized by integer Chern-Simons numbers, $N_{\mathrm{CS}} = n \in \mathbb{Z}$. Our numerical calculations confirm that for $q = 0$, we obtain the trivial vacuum with $N_{\mathrm{CS}} = 0$. For $q = 2\pi$, we obtain the vacuum with $N_{\mathrm{CS}} = 1$. For $q = 4\pi$, we obtain the vacuum with $N_{\mathrm{CS}} = 2$. This follows directly from Eq.~(\ref{eq:Nradial}), as the integral term vanishes for vacuum configurations and $N = (q - \sin q)/(2\pi)$, which equals 0, 1, and 2 for $q = 0$, $2\pi$, and $4\pi$, respectively. These different vacua, labeled by consecutive integers, are separated by energy barriers in the configuration space. The sphaleron represents the static solution at the top of this barrier. By symmetry considerations, the sphaleron between vacua with $N_{\mathrm{CS}} = n$ and $N_{\mathrm{CS}} = n+1$ has $N_{\mathrm{CS}} = n + \frac{1}{2}$.
\begin{equation}
N_{\mathrm{CS}} = n + \tfrac{1}{2} \, .
\label{eq:sphaleronNCS}
\end{equation}
Our numerical calculations verify that: For $q = \pi$, we obtain the sphaleron with $N_{\mathrm{CS}} = \frac{1}{2}$,  For $q = 3\pi$, we obtain the sphaleron with $N_{\mathrm{CS}} = \frac{3}{2}$.

Again, this follows from Eq.~(\ref{eq:Nradial}), as $(q - \sin q)/(2\pi) = 1/2$ for $q = \pi$ and $3/2$ for $q = 3\pi$. The sphaleron with $N_{\mathrm{CS}} = \frac{1}{2}$ is the most relevant for Standard Model phenomenology, representing the minimal-energy barrier configuration between the trivial vacuum and the vacuum with unit winding number.

%%%%%%%%%%%%%%%%%%%%%%%%%%%%%%%%%%%%%%%%%%%%%%%%%%%%%%%%%%%%%%%%%%%%%
\subsection{Lagrange Multipliers for Static Configurations}
%%%%%%%%%%%%%%%%%%%%%%%%%%%%%%%%%%%%%%%%%%%%%%%%%%%%%%%%%%%%%%%%%%%%%
To find static configurations numerically and explore the minimum-energy path connecting adjacent vacua, we use a constrained minimization approach following Ref.~\cite{Akiba:1988ay}. We minimize the static energy functional $E_{\rm static}$ subject to the constraint of fixed Chern-Simons number.

We introduce a Lagrange multiplier $\kappa$ and define the modified functional:
\begin{equation}
\tilde{E}[\psi] = E_{\rm static}[\psi] + \kappa \, N_{\rm CS}[\psi] \, ,
\label{eq:modifiedE}
\end{equation}
where $\psi$ represents the set of profile functions $\{f_A(r), f_B(r), H(r), K(r)\}$. The parameter $\kappa$ serves as a Lagrange multiplier enforcing the Chern-Simons number constraint. This leads to $ d \tilde{E}/d \kappa = N_{\rm CS}$. This approach modifies the equations of motion for the gauge field components. The modified versions of Eqs.~(\ref{eq:fa2}) and~(\ref{eq:fb2}) become:
\begin{align}
&f_A'' - \frac{1}{r^2} \left(f_A^2 + f_B^2 - 1 \right) f_A~\nonumber \\
&- m_W^2 \left[(H^2+ K^2)f_A + K^2 - H^2 \right] + \zeta m_W f_B' = 0 \, , \\
& f_B'' - \frac{1}{r^2} \left(f_A^2 + f_B^2 - 1 \right) f_B~\nonumber \\
&- m_W^2 \left[(H^2+ K^2)f_B -2HK\right] - \zeta m_W f_A' = 0  \, ,
\end{align}
where $\zeta = \alpha_W \kappa/(2 \pi m_W)$ and parameterizes the deviation from the unconstrained equations. The Higgs field equations remain unchanged in form but are coupled to the modified gauge field equations through the solutions for $f_A$ and $f_B$.

By systematically varying $\zeta$ from $0$ to some maximum value and back to $0$, we can trace the minimum-energy path connecting the vacuum at $N_{\rm CS} = 0$ through the sphaleron at $N_{\rm CS} = \frac{1}{2}$ to the vacuum at $N_{\rm CS} = 1$. The energy along this path forms a potential barrier whose height determines the tunneling probability at zero temperature and the thermal activation rate at finite temperature~\cite{Kuzmin:1985mm}.

The boundary conditions for the profile functions must be modified to account for the Lagrange multiplier term. Near the origin, $r \simeq 0$, the series expansions modify the boundary condition $f_B$:
\begin{equation}
    f_B \; \simeq \; -\frac{1}{3} \left(\gamma_H \gamma_K - \zeta \gamma_A \right)(m_W r)^3 \, ,
\end{equation}
and at $r = \infty$, we find
\begin{align}
f_A &= \cos q + \mathrm{Re} \left[ (\delta_A \cos q - \delta_B \sin q) \exp(-\beta r) \right] \, , \\
f_B &= \sin q + \mathrm{Re}  \left[ (\delta_A \sin q + \delta_B \cos q) \exp(-\beta r) \right] \, , \\
H &= \cos\frac{q}{2} \left[ 1 + \frac{\delta_H}{m_W r} \exp(-m_H r) \right] ~\nonumber\\ 
&+ \sin\frac{q}{2} \, \mathrm{Re}  \left[ \frac{\delta_B}{\beta^2 r^2} \exp(-\beta r) \right] \, , \\
K &= \sin\frac{q}{2} \left[ 1 + \frac{\delta_H}{m_W r} \exp(-m_H r) \right] ~\nonumber \\
&- \cos\frac{q}{2} \, \mathrm{Re}  \left[ \frac{\delta_B}{\beta^2 r^2} \exp(-\beta r) \right] \, ,
\end{align}
where $\delta_H$ is real, and $\delta_A$, $\delta_B$, and $\beta$ are complex numbers with the condition $\delta_B = m_W \beta \zeta \delta_A/(m_W^2 - \beta^2)$, and the real parameter $\beta$ satisfies the constraint $(\beta^2 - m_W^2)^2 = m_W^2 \beta^2 \zeta^2$.
\begin{figure*}[ht!]
    \centering
    
    % --- Top row ---
    \begin{subfigure}[b]{0.49\textwidth}
        \centering
        \includegraphics[width=\textwidth]{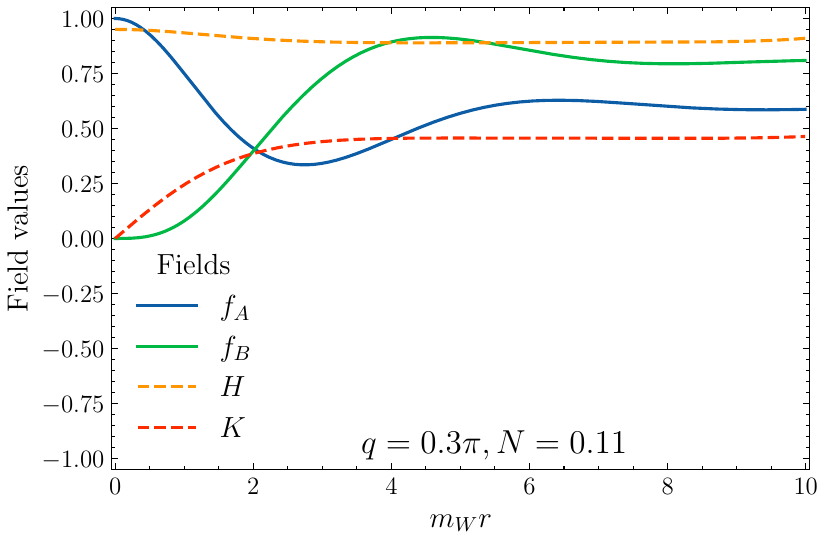}
    \end{subfigure}
    \hfill
    \begin{subfigure}[b]{0.49\textwidth}
        \centering
        \includegraphics[width=\textwidth]{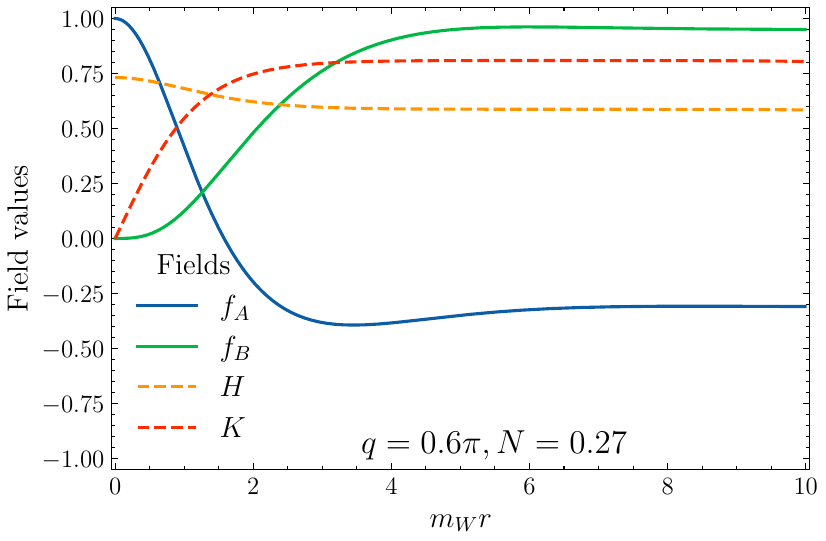}
    \end{subfigure}
    
    \vspace{0.5cm}
    
    % --- Bottom row ---
    \begin{subfigure}[b]{0.49\textwidth}
        \centering
        \includegraphics[width=\textwidth]{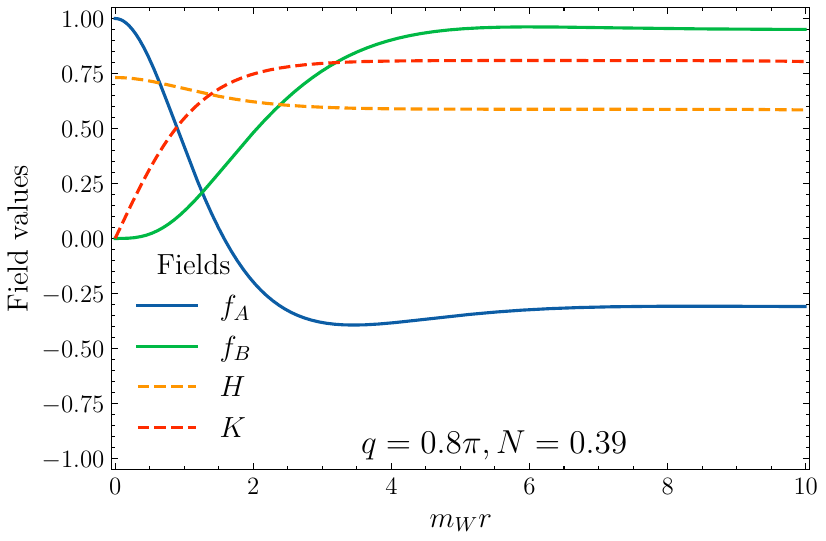}
    \end{subfigure}
    \hfill
    \begin{subfigure}[b]{0.49\textwidth}
        \centering
        \includegraphics[width=\textwidth]{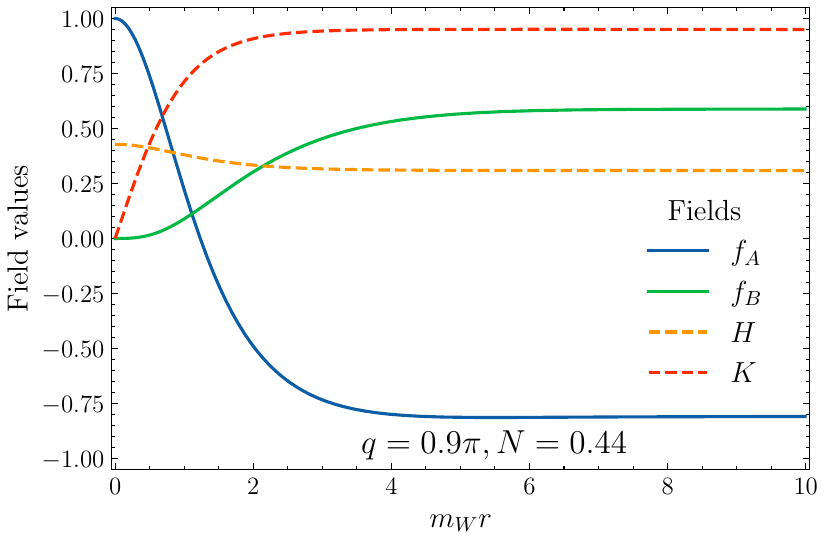}
    \end{subfigure}
    
    \caption{Field profiles of the sphaleron configuration and vacuum solutions for different values of the parameter $q$, which parametrizes a family of field configurations interpolating between topologically distinct vacua in $SU(2)_{\rm W}$ gauge theory. Each panel shows the radial dependence of the gauge field functions $f_A$, $f_B$, and the scalar fields $H$, $K$ as functions of $m_W r$. The topological number associated with each configuration is given by $N$. The panels correspond to intermediate configurations between vacua and sphalerons, with increasing $q$ leading to larger topological charge $N$ and energy. This sequence demonstrates how increasing $q$ leads to configurations with progressively higher topological charge and energy.}
    \label{fig:solslagrange}
\end{figure*}

This constrained minimization approach allows us to systematically compute the minimum-energy path connecting topologically distinct vacua through the sphaleron. The resulting energy barrier profile is crucial for determining baryon-number-violating transition rates in the early universe~\cite{Kuzmin:1985mm,Shaposhnikov:1987tw} and potentially in high-energy collisions~\cite{Tye:2015tva,Ellis:2016ast}.

The approach to studying sphalerons outlined above follows the constrained minimization method of Akiba, Kikuchi, and Yanagida (AKY)~\cite{Akiba:1988ay}, which differs methodologically from the original approach of Manton~\cite{Manton:1983nd}. The AKY approach, which we adopt here, directly constrains the topological charge through a Lagrange multiplier, allowing for a systematic exploration of the minimum-energy path connecting adjacent vacua. In contrast, Manton's approach uses a specific parametrization that interpolates between vacua. These different methods of constructing the path between vacua represent complementary perspectives on the sphaleron transition. The AKY approach has the advantage of directly targeting configurations with specific Chern-Simons numbers, which is particularly useful for mapping out the detailed energy landscape between topological sectors. This systematic exploration of the minimum-energy path is especially valuable when studying potential sphaleron production mechanisms in both cosmological and high-energy collision contexts.

Fig.~\ref{fig:solslagrange} illustrates the radial profiles of the field functions for various configurations along the path connecting multiple topological sectors. As we vary the parameter $q$ from 0 to $\pi$, we observe the systematic evolution of the field profiles corresponding to vacua at $q = 0$, (with $N_{\rm CS} = 0$) and sphalerons at $q = \pi$ (with $N_{\rm CS} = 1/2$). The gauge field components $f_A(r)$ and $f_B(r)$ approach pure gauge configurations at large distances, while the Higgs field components $H(r)$ and $K(r)$ approach their vacuum expectation values with phase rotations determined by $q$. These profiles clearly demonstrate the topological distinction between different sectors and the barrier configurations (sphalerons) that separate them. The energy of these configurations increases with $q$ up to $q = 2\pi$ and then follows a periodic pattern, reflecting the cyclic nature of the vacuum structure in $SU(2)$ gauge theory with a Higgs field.

%%%%%%%%%%%%%%%%%%%%%%%%%%%%%%%%%%%%%%%%%%%%%%%%%%%%%%%%%%%%%%%%%%%%%
\section{Eigenvalue Equations and Unstable Modes}
\label{sec:eigenvalueequations}
%%%%%%%%%%%%%%%%%%%%%%%%%%%%%%%%%%%%%%%%%%%%%%%%%%%%%%%%%%%%%%%%%%%%%

%%%%%%%%%%%%%%%%%%%%%%%%%%%%%%%%%%%%%%%%%%%%%%%%%%%%%%%%%%%%%%%%%%%%%
\subsection{Coupled Channels}
%%%%%%%%%%%%%%%%%%%%%%%%%%%%%%%%%%%%%%%%%%%%%%%%%%%%%%%%%%%%%%%%%%%%%
We now proceed to analyze the dynamical stability of the sphaleron by computing its unstable modes. These results are fundamentally important for understanding the decay dynamics of sphalerons and are applied in our companion paper on lattice studies of sphaleron decay~\cite{MatchevVerner2025b}.\footnote{For the publicly available sphaleron lattice decay code on GitHub, see \url{https://github.com/sarunasverner/sphaleron-decay}.}

The sphaleron solution in the radial gauge can be expressed in the Manton parametrization~\cite{Manton:1983nd,Klinkhamer:1984di} as
\begin{align}
    f_A(r, t) \; = \; 1 - 2 f(r) \, , \\
    K(r, t) \; = \; h(r) \, , \\
    f_B(r,t) \; = \; f_C(r,t) \; = \; H(r,t) \; = \; G(r, t) \; = \; 0 \, ,
\end{align}
where $f(r)$ and $h(r)$ are radial profile functions that satisfy the boundary conditions:
\begin{align}
    f(0) &= 0 \,, \quad f(\infty) = 1 \, , \\
    h(0) &= 0 \, , \quad h(\infty) = 1 \, .
\end{align}

\begin{figure}[t!]
    \centering
    \includegraphics[width=\linewidth]{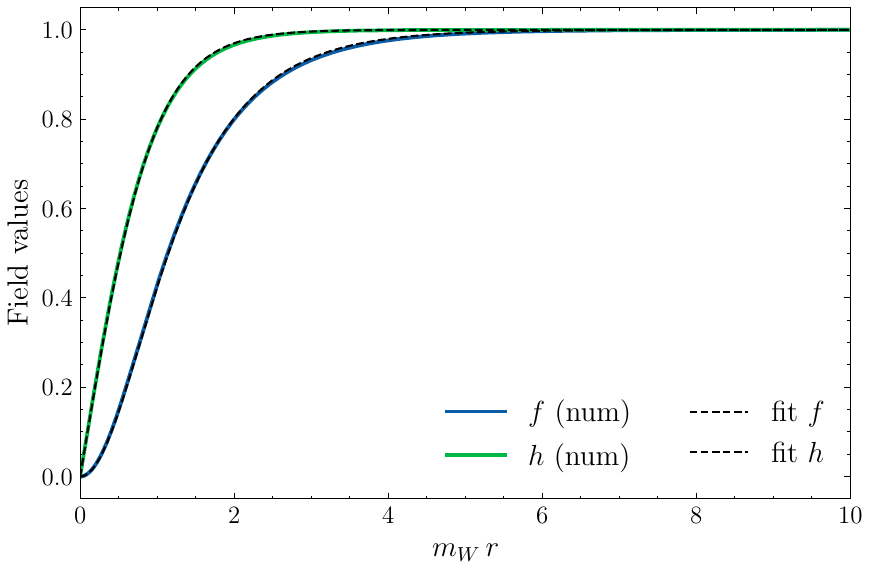}
    \caption{Radial profile functions for the sphaleron solution. Solid lines show the numerically computed profiles obtained from our constrained minimization approach, while dashed lines show the analytical approximations using hyperbolic functions as proposed in Ref.~\cite{Tye:2015tva}. The agreement validates both the numerical and analytical approaches.}
    \label{fig:fandhfits}
\end{figure}

As shown by Tye and Wong~\cite{Tye:2015tva}, these profile functions can be accurately approximated by hyperbolic functions:
\begin{equation}
    f(r) = 1 - \sech(1.15m_Wr)\,, \quad h(r) = \tanh(1.05 m_W r) \, .
\end{equation}
Fig.~\ref{fig:fandhfits} compares our numerically computed profiles with these analytical approximations, showing excellent agreement.

To analyze the stability of the sphaleron solution, we introduce small perturbations around the static configuration and study their time evolution. Following Ref.~\cite{Akiba:1989xu}, we define fluctuation fields $\eta_i(r,t)$ (with $i = A, B, C, H, K$) as:
\begin{align}
f_A(r,t) &= 1 - 2f(r) + r\,\eta_A(r,t) \, ,  \\
f_B(r,t) &= r\,\eta_B(r,t) \, , \\
f_C(r,t) &= \sqrt{2}\,\eta_C(r,t) \, ,  \\
H(r,t)   &= \frac{1}{\sqrt{2}m_W}\,\eta_H(r,t) \, ,  \\
K(r,t)   &= h(r) + \frac{1}{\sqrt{2}m_W}\,\eta_K(r,t) \, .
\end{align}
The factors in these definitions are chosen to simplify the kinetic energy term in the Hamiltonian. With the time component of the gauge field set to zero, $G(r,t) = 0$, the kinetic energy contribution from Eq.~(\ref{eq:kinen}) becomes:
\begin{equation} 
    E_{\rm kin} \; = \; \frac{4 \pi}{g^2} \int_0^{\infty} r^2 dr \left(\dot{\eta}_A^2 + \dot{\eta}_B^2 + \dot{\eta}_C^2 + \dot{\eta}_H^2 + \dot{\eta}_K^2 \right) \, ,
    \label{eq:fluckin}
\end{equation}
where the dot denotes a time derivative. This form ensures that the kinetic term has the standard canonical structure.

To analyze the stability of the sphaleron, we expand the static energy functional around the sphaleron solution to second order in the fluctuation fields:
\begin{equation}
\begin{aligned}
    &E_{\rm static} = E_{\rm sph} \\
    &+ \frac{1}{2}
         \int_{0}^{\infty} dr
         \int_{0}^{\infty} dr'\;
         \eta_i(r)\,
         \mathcal{H}_{ij}(r,r')\,
         \eta_j(r')
       + \cdots \, ,
       \label{eq:Hexp}
\end{aligned}       
\end{equation}
where 
\begin{equation}
     \mathcal{H}_{ij}(r,r') = 
         \left. \frac{\delta^{2} E_{\text{static}}}
              {\delta\eta_i(r)\,\delta\eta_j(r')}
       \right|_{\rm sph} \, ,
       \label{eq:Hessian}
\end{equation}
is the kernel of second functional derivatives (the Hessian) evaluated at the sphaleron background. This Hessian acts as the ``mass matrix" in the linearized fluctuation problem, and its eigenspectrum determines the stability of the sphaleron.

A crucial property of the Hessian, first noted by Ref.~\cite{Akiba:1989xu}, is that it decomposes into two independent sectors due to the symmetries of the spherically symmetric ansatz: the two‑channel
sector $\boldsymbol\eta(r) = (\eta_A,\eta_K)^{T}$ and the
three‑channel sector $\boldsymbol\eta(r) = (\eta_B,\eta_C,\eta_H)^{T}$. This decomposition significantly simplifies the eigenvalue analysis, allowing us to treat these two sectors independently.

In this case, the operator form for the two coupled channel is given by
\begin{widetext}
\begin{equation}
     \mathcal L_2(r)\,\boldsymbol\eta(r) = \omega^{2}\,\boldsymbol\eta(r)  \qquad\text{with}\qquad
\mathcal L_2(r)
  = \begin{pmatrix}
        -\PrOp^{2} + \dfrac{2}{r^{2}} + m_W^{2} + U_{AA}(r) &
        \dfrac{1}{\sqrt{2}}\,U_{AK}(r) \\
        \dfrac{1}{\sqrt{2}}\,U_{AK}(r) &
          -\PrOp^{2} + m_H^{2} + U_{KK}(r)
      \end{pmatrix} \, ,
\end{equation}
where $ \PrOp \;=\; \frac1r\dr\,r$ and the radial potentials are given by
\begin{align}
  U_{AA}(r) &= \frac{3}{r^{2}}\!\Bigl[(1-2f)^2-1\Bigr]
              + m_W^{2}\bigl[h^{2}-1\bigr] \, , \\
  U_{AK}(r) &= \frac{4m_W}{r}\,h\,(1-f) \, , \\
  U_{KK}(r) &= \frac{2}{r^{2}}(1-f)^{2}
              + \frac{3}{2}m_H^{2}\bigl[h^{2}-1\bigr] \, .
\end{align}

For the three-coupled channels, we find 
\begin{equation}
\mathcal L_3(r) \,\boldsymbol\eta(r)=\omega^{2}\,\boldsymbol\eta(r) \, ,
\end{equation}
with block entries
\begin{equation}
\mathcal L_3(r)= 
\begin{pmatrix}
  -\PrOp^{2} + m_W^{2}+U_{BB} &
  \sqrt2\!\left(-\dfrac1r\PrOp+\dfrac1{r^{2}}+U_{BC}\right) &
  -\sqrt2\!\left(\dfrac{m_W}{r}+U_{BH}\right) \\
  \sqrt2\!\left(\dfrac1r\PrOp+U_{CB}\right) &
  2\left(\dfrac{1}{r^{2}}+\dfrac12 m_W^{2}+U_{CC} \right) &
  m_W\PrOp-\dfrac{m_W}{r}+U_{CH} \\
  -\sqrt2\!\left(\dfrac{m_W}{r}+U_{HB}\right) &
  -m_W\PrOp-\dfrac{m_W}{r}+U_{HC} &
  -\PrOp^{2} + \dfrac{2}{r^{2}}+U_{HH}
 \end{pmatrix} \, ,
\end{equation}
where
\begin{align}
  U_{BB}(r) &= \frac{1}{r^{2}}\Bigl[(1-2f(r))^{2}-1\Bigr]
              + m_{W}^{2}\bigl[h(r)^{2}-1\bigr] \, , \\
  U_{CC}(r) &= \frac{1}{r^{2}}\Bigl[(1-2f(r))^{2}-1\Bigr]
              + \frac12\,m_{W}^{2}\bigl[h(r)^{2}-1\bigr] \, , \\
  U_{HH}(r) &= -\frac{2}{r^{2}}\Bigl[1-f(r)^{2}\Bigr]
              + \frac12\,m_{H}^{2}\bigl[h(r)^{2}-1\bigr] \, ,\\
  U_{BC}(r) &= \frac{2}{r}\!\left[-2f'(r)
                 +(1-f(r))\frac{d}{dr}\right] \, , \\
  U_{CB}(r) &= -\frac{2}{r}f'(r)
              -\frac{2}{r}(1-f(r))\!\left[\frac{1}{r}+\frac{d}{dr}\right] \, , \\
  U_{BH}(r) &= U_{HB}(r)=\frac{m_{W}}{r}\bigl[h(r)-1\bigr] \, , \\
  U_{CH}(r) &= m_{W}\bigl[h(r)-1\bigr]\frac{d}{dr}
              -m_{W}h'(r) \, , \\
  U_{HC}(r) &= -2m_{W}h'(r)
              -m_{W}\bigl[h(r)-1\bigr]\!\left[\frac{2}{r}+\frac{d}{dr}\right] \, .
\end{align}
\end{widetext}
These operators determine the normal modes of oscillation around the sphaleron. The eigenvalue problem for the normal modes is given by:
\begin{equation}
\mathcal L_N(r)\,\boldsymbol\eta_n(r)=\omega_n^{2}\,\boldsymbol\eta_n(r),
\qquad N=2,3 \, ,
\end{equation}
where $\omega^2$ are the eigenvalues. Negative eigenvalues ($\omega^2 < 0$) correspond to unstable modes, while positive eigenvalues ($\omega^2 > 0$) correspond to stable oscillatory modes. The lowest eigenfrequency $\omega_-$ identifies the single unstable (tachyonic) direction of the sphaleron.

All eigenfrequencies $\omega^{2}$ are guaranteed to be real because the radial momentum operator is anti-Hermitian, $\hat{P}_{r}^{\dagger} = -\hat{P}_{r}$, and the mixed potential terms obey $U_{ij}^{\dagger} = U_{ji}$. This ensures that the Hessian operator is Hermitian.

%%%%%%%%%%%%%%%%%%%%%%%%%%%%%%%%%%%%%%%%%%%%%%%%%%%%%%%%%%%%%%%%%%%%%
\subsection{The Unstable Mode}
%%%%%%%%%%%%%%%%%%%%%%%%%%%%%%%%%%%%%%%%%%%%%%%%%%%%%%%%%%%%%%%%%%%%%
A crucial aspect of the stability analysis is the proper treatment of gauge degrees of freedom. The three-channel sector ($\eta_B$, $\eta_C$, and $\eta_H$) contains an infinite number of gauge zero modes, which must be excluded from the physical energy spectrum. These zero modes represent gauge transformations of the sphaleron configuration and do not correspond to physical fluctuations.

Following Ref.~\cite{Akiba:1989xu}, we implement a gauge-fixing condition:
\begin{equation}
    (r^2 \eta_C)' \; = \; \sqrt{2}(1-2f) r \eta_B - m_W r^2 h \eta_H \, ,
    \label{eq:gaugefixing}
\end{equation}
where the prime denotes differentiation with respect to $r$. This condition ensures that only physical modes are considered in our analysis.

It is worth noting that diagonal entries in the potential matrices exhibit singular behaviors near the origin. Specifically, $U_{HH}$ contains an attractive $1/r^{2}$ singularity, whereas $U_{KK}$ carries a repulsive $1/r^{2}$ term. No other channel is singular near the origin, since $f(r) \sim r^{2}$ as $r \to 0$. These singularities are pure gauge artifacts that would disappear after transforming to an appropriate background gauge. However, we refrain from applying such a transformation here, as it would mix states of even and odd parity and thus obscure our channel decomposition.

For a well-defined eigenvalue problem, we must specify appropriate boundary conditions for the fluctuation fields. At the origin ($r \to 0$), regularity requires:
\begin{align}
 r \, \eta_i(r)|_{r = 0} \; = \; 0 \, .
\end{align}
In general, at large distances ($r \to \infty$), the fluctuations must be normalizable, which requires:
\begin{align}
\eta_i(r) \sim \frac{e^{-k_i r}}{r} \quad \text{as} \quad r \to \infty \, ,
\end{align}
where $k_i$ depends on the asymptotic form of the potential terms and the eigenvalue $\omega^2$. For stable modes ($\omega^2 > 0$), we have $k_i = \sqrt{m_i^2 - \omega^2}$, where $m_i$ is the effective mass of the corresponding field. For the unstable mode ($\omega_-^2 < 0$), we have $k_i = \sqrt{m_i^2 + |\omega_-^2|}$. We note that $\eta_C$ is not subject to the boundary condition at $r = \infty$, since the terms $\eta_C'$ do not appear in Eq.~(\ref{eq:Hexp}).

To solve the eigenvalue problem numerically, we discretize the radial coordinate on a logarithmic grid that extends from $r_{\text{min}} = 10^{-4} \, m_W^{-1}$ to $r_{\text{max}} = 20 \, m_W^{-1}$ with $1000$ grid points. This provides sufficient resolution near the origin while capturing the asymptotic behavior at large distances. We implement the gauge-fixing condition~(\ref{eq:gaugefixing}) by eliminating $\eta_C$ in favor of $\eta_B$ and $\eta_H$. The resulting eigenvalue problem is solved using a combination of finite-difference methods and sparse matrix diagonalization techniques.

Our numerical analysis reveals that the sphaleron possesses exactly one unstable mode, with eigenvalue:
\begin{equation}
    \omega_{-}^2 \simeq -2.7 \, m_W^2 \, ,
\end{equation}
corresponding to an imaginary frequency $\omega_{-} \simeq 1.64 i\,m_W$. Previously, Ref.~\cite{Akiba:1989xu} found $ \omega_{-}^2 \simeq -2.3 \, m_W^2$ assuming that $m_W = m_H$.

The unstable mode resides in the three-channel sector, coupling the gauge field fluctuation $\eta_B$ and $\eta_C$ and the Higgs field fluctuation $\eta_H$. This negative eigenvalue confirms that the sphaleron is indeed a saddle point of the energy functional with exactly one unstable direction, as required for a barrier configuration between adjacent topological vacua.

\begin{figure}
    \centering
    \includegraphics[width=1.0\linewidth]{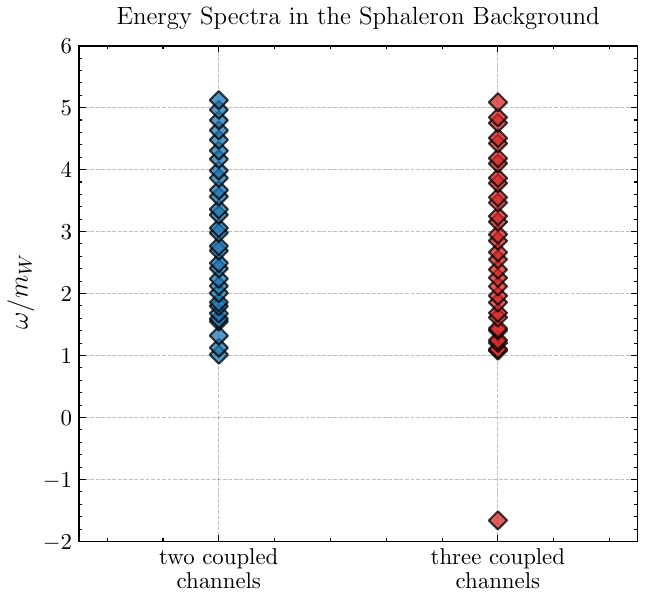}
    \caption{Low-lying eigenvalue spectrum of fluctuations around the sphaleron. Left: Two-channel sector showing only stable modes after gauge fixing. Right: Three-channel sector showing the unstable mode with $\omega^2 < 0$ and several stable modes. The eigenvalues are given in units of $m_W$, so that $\omega_-$ is plotted as ${\rm sign}({\omega_-^2})\sqrt{|\omega_-^2|}$.}
    \label{fig:energysepctra}
\end{figure}

Fig.~\ref{fig:energysepctra} shows the low-lying energy spectrum for both the two-channel and three-channel sectors. The spectrum includes the unstable mode with $\omega_-^2 < 0$ in the three-channel sector, as well as several stable modes with $\omega^2 > 0$ in both sectors. The lowest non-zero eigenvalues in the three-channel sector correspond to gauge rotation modes, while higher eigenvalues represent physical oscillations around the sphaleron configuration.

The existence and properties of the unstable mode are crucial for understanding sphaleron-mediated baryon number violation. The eigenvalue $\omega_{-}^2$ determines the characteristic time scale for sphaleron decay, $\tau_{\text{decay}} \simeq 1/|\omega_{-}| \simeq 0.61 m_W^{-1}$, which corresponds to approximately $5 \times 10^{-27}$ seconds.

At zero temperature baryon-number violation proceeds through
instanton tunnelling and is suppressed by the Euclidean action
$S_E = 4\pi/\alpha_W$.  With the physical coupling
$\alpha_W \simeq 0.034$ one finds
$\exp(-S_E)\simeq   \exp(-370)\sim10^{-161}$, making the process utterly negligible~\cite{tHooft:1976rip}. Near the electroweak crossover ($T_c\simeq159\;\text{GeV}$~\cite{DOnofrio:2014rug})
thermal activation over the sphaleron barrier dominates.
In the broken phase the leading weak-coupling result
\cite{Arnold:1987mh, Carson:1990jm,Arnold:1996dy} is
\begin{equation}
\frac{\Gamma}{V}
   \; \propto \;
  \frac{\omega_{-}}{2\pi} \left(\frac{\alpha_W T}{4\pi} \right)^3
   \exp\!\Bigl[-\frac{E_{\rm sph}(T)}{T}\Bigr] \, .
\label{eq:sphrate}
\end{equation}
The explicit dependence on $\omega_-$ shows why an accurate value
of the negative mode is essential for quantitative baryogenesis
computations.

In our companion paper~\cite{MatchevVerner2025b}, we use the unstable mode calculated here to initialize real-time lattice simulations of sphaleron decay. By perturbing the sphaleron configuration along the unstable direction and evolving the full non-linear field equations, we can directly study the dynamical process of baryon number violation, including non-perturbative effects that are difficult to capture in other approaches.

%%%%%%%%%%%%%%%%%%%%%%%%%%%%%%%%%%%%%%%%%%%%%%%%%%%%%%%%%%%%%%%%%%%%%
\subsection{Analytical Formulation for the Three-Channel Sector}
%%%%%%%%%%%%%%%%%%%%%%%%%%%%%%%%%%%%%%%%%%%%%%%%%%%%%%%%%%%%%%%%%%%%%
The eigenvalue analysis in the previous sections reveals that the sphaleron possesses exactly one unstable mode with eigenvalue $\omega_{-}^2 \simeq -2.7\,m_W^2$. Importantly, this mode resides in the three-channel sector (coupling $\eta_B$, $\eta_C$ and $\eta_H$). Therefore, for a comprehensive understanding of the sphaleron dynamics, the analysis of this sector is essential.

Following Akiba et al.~\cite{Akiba:1989xu}, we can analyze the three-channel sector by focusing on the coupled differential equations for $\eta_B$, $\eta_C$, and $\eta_H$. After implementing the gauge-fixing condition, the linearized equations of motion around the sphaleron background take the form:
\begin{align}
\label{eq:BB}
&\eta_B'' + \frac{2}{r}\eta_B' + \Bigl[\omega^{2}-m_{W}^{2}h^{2}-\frac{2}{r^{2}}+\frac{12}{r^{2}}f(1-f)\Bigr]\eta_B \\
\label{eq:CC}
&+ \frac{2\sqrt{2}}{r^{3}}\bigl(1-2f+2r\,f'\bigr)\eta_C + \frac{2\sqrt{2}}{r^{2}}m_{W}(1-f)h\,\eta_C = 0 \,,~\nonumber \\
&r\,\eta_C' + 2\,\eta_C - \sqrt{2}(1-2f)\eta_B + m_{W}h\,r\,\eta_H = 0 \, , \\
\label{eq:HH}
&\eta_H'' + \frac{2}{r}\eta_H' + \Bigl[\omega^{2}-m_{W}^{2}h^{2}+\tfrac12 m_{H}^{2}(1-h^{2})-\frac{2}{r^{2}}f^{2}\Bigr]\eta_H \nonumber\\
&+ \frac{2\sqrt{2}}{r^{2}}m_{W}(1-f)h\,\eta_B + 2m_{W}h'\,\eta_C = 0 \, . 
\end{align}
These equations clarify how the fluctuations couple to the background sphaleron configuration through the profile functions $f(r)$ and $h(r)$. We note that Eq.~(\ref{eq:CC}) is precisely the gauge-fixing condition Eq.~(\ref{eq:gaugefixing}) rearranged to express $\eta_C$ in terms of the other fluctuations. This approach eliminates the gauge zero modes while preserving the coupled structure of the physical modes.

The boundary conditions for these equations follow from our earlier analysis of regularity at the origin and normalizability at infinity. For numerical purposes, we implement these as:

\begin{align}
r\eta_i(r)|_{r = 0} &= 0 \, , \quad \text{(regularity at the origin)} \, , \\
\eta_i(r)|_{r = R} &= 0 \, . \quad \text{(normalizability at large $R$)} \, ,
\end{align}
where $R$ is a sufficiently large numerical cutoff, typically $R \gtrsim 10 m_W^{-1}$ in our computations.

To solve the coupled system Eqs.~(\ref{eq:BB})-(\ref{eq:HH}), we employ a shooting method with fourth-order Runge-Kutta integration, matching the solutions at an intermediate point to ensure global consistency. Our approach generalizes the method developed by Ref.~\cite{Akiba:1989xu}, but incorporates the $m_H = 125.1$ GeV and $m_W = 80.4$ GeV rather than the approximation $m_H = m_W$ used in earlier works. Our numerical analysis confirms that the sphaleron possesses exactly one unstable mode with eigenvalue $\omega_{-}^2 \simeq -2.7\,m_W^2$, in agreement with the results in the previous section. This unstable mode resides in the three-channel ($\eta_B$-$\eta_C$-$\eta_H$) sector, which is precisely the direction in configuration space that connects adjacent topological vacua. 
\begin{figure}
    \centering
    \includegraphics[width=1\linewidth]{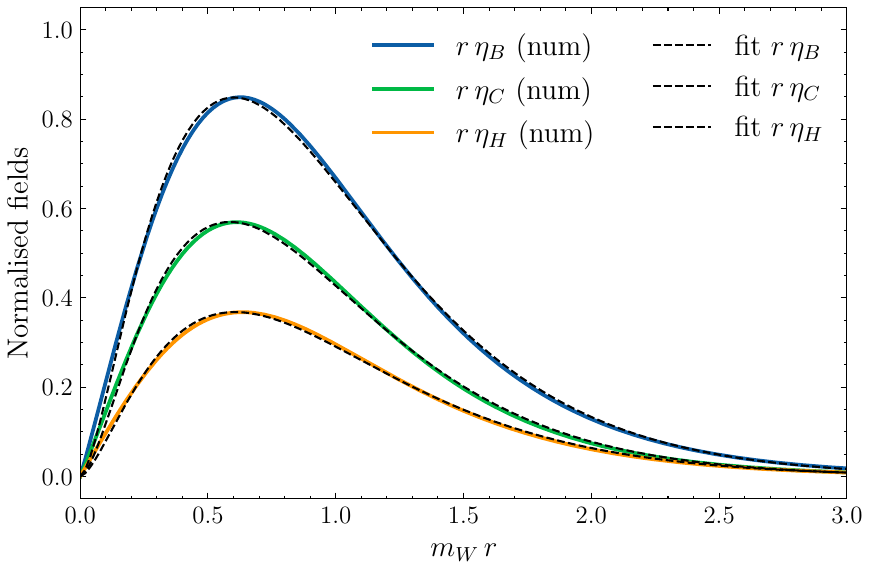}
    \caption{Normalized profiles of the fluctuation modes $\eta_B$, $\eta_C$, and $\eta_H$ in the three-channel sector for the lowest stable eigenfrequency. These profiles illustrate the coupled nature of gauge and Higgs field fluctuations around the sphaleron.}
    \label{fig:normprofiles}
\end{figure}

The radial dependence of these mode profiles can be accurately approximated by the following analytical expressions:
\begin{align}
    \eta_B(r) &= 10.22\, e^{-2.74\,m_W r} (m_W r)^{0.65} \, , \\
    \eta_C(r) &= 7.48\, e^{-2.86\,m_W r} (m_W r)^{0.68} \, , \\
    \eta_H(r) &= 4.00\, e^{-2.62\,m_W r} (m_W r)^{0.59} \, .
\end{align}
The wavefunctions are normalized according to:
\begin{equation}
    \label{eq:normalizationgen}
    \int_0^R dr \, r^2 \left[\eta^2_B(r) + \eta^2_C(r) + \eta^2_H(r) \right] \; = \; m_W^{-1} \, .
\end{equation}
Fig.~\ref{fig:normprofiles} shows the normalized profiles of the fluctuation modes in the unstable direction. The profiles reveal important physical characteristics of fluctuations around the sphaleron. The gauge field fluctuations ($\eta_B$ and $\eta_C$) are more pronounced at shorter distances compared to the Higgs field fluctuation ($\eta_H$), reflecting the different coupling strengths to the background sphaleron configuration. All fluctuations exhibit exponential localization with characteristic length scales of order $m_W^{-1}$, consistent with the typical size of the sphaleron.

These analytical fits are particularly useful for initializing lattice simulations of sphaleron dynamics, as they provide a compact and accurate representation of the eigenmodes without requiring the full numerical solution of the differential equations at each lattice site. In our companion paper~\cite{MatchevVerner2025b}, we employ these mode functions to study real-time evolution of perturbed sphaleron configurations, allowing us to directly observe the non-linear dynamics of baryon number violation.

%%%%%%%%%%%%%%%%%%%%%%%%%%%%%%%%%%%%%%%%%%%%%%%%%%%%%%%%%%%%%%%%%%%%%
\section{One-dimensional Potential Barrier}
\label{sec:1dbarrier}
%%%%%%%%%%%%%%%%%%%%%%%%%%%%%%%%%%%%%%%%%%%%%%%%%%%%%%%%%%%%%%%%%%%%%
Having characterized the static sphaleron solution and analyzed its stability properties, we now develop an effective model to describe the dynamics of sphaleron-mediated transitions between topologically distinct vacua. This model provides valuable insights into baryon number violation mechanisms and serves as the foundation for numerical simulations of sphaleron decay. Our stability analysis revealed that the sphaleron possesses exactly one unstable mode with eigenvalue $\omega_{-}^2 \simeq -2.7\,m_W^2$. This negative eigenvalue defines the unique direction in field configuration space along which the system can evolve away from the saddle point, potentially transitioning between topologically distinct vacua. Following the approach pioneered by Refs.~\cite{Ringwald:1987ej, Kripfganz:1989hu}, we project the full field dynamics onto this unstable direction using a collective coordinate formalism. This approach effectively reduces the infinite-dimensional field theory to a one-dimensional quantum mechanical system, capturing the essential physics of topological transitions while remaining computationally tractable.

Based on our detailed stability analysis, we parameterize field configurations in the vicinity of the sphaleron using the unstable mode. The complete field configuration is given by:
\begin{align}
    f_A(r, t) \; = \; 1 - 2 f(r)  \, , \\
    K(r, t) \; = \; h(r) \, , \\
    f_B(r,t) \; = \; \kappa(t) r \eta_B(r) \, , \\
    f_C(r,t) \; = \; \sqrt{2} \kappa(t) \eta_C(r) \, , \\
    H(r, t) \; = \; \frac{1}{\sqrt{2}m_W}\kappa(t) \eta_H(r) \, .
\end{align}
Here, $f(r)$ and $h(r)$ are the static sphaleron profile functions described in Section~\ref{sec:sphstaticminimum} while $\eta_A(r)$, $\eta_B(r)$, $\eta_C(r)$, $\eta_H(r)$, and $\eta_K(r)$ are the components of the fluctuation fields. The time-dependent collective coordinate $\kappa(t)$ measures the amplitude of the fluctuation along the unstable direction, with $\kappa = 0$ corresponding to the exact sphaleron configuration.

The mode functions are normalized according to Eq.~(\ref{eq:normalizationgen}). This normalization ensures that the effective mass parameter in our one-dimensional model has the appropriate dimensions and scaling with the gauge coupling.

\begin{figure}
    \centering
    \includegraphics[width=0.9\linewidth]{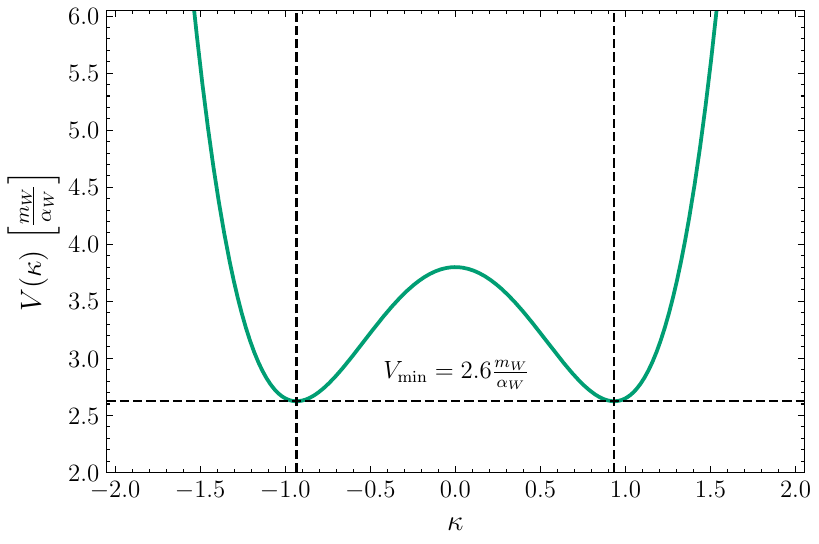}
    \caption{Effective potential $V(\kappa)$ for the sphaleron transition as a function of the collective coordinate $\kappa$. The potential exhibits a double-well structure with a barrier height of $3.8m_W/\alpha_W \simeq 9.1 \rm{TeV}$ at $\kappa = 0$ (the sphaleron configuration) and minima at $\kappa \simeq \pm 0.9$ (the adjacent topological vacua). The shape of this potential determines the quantum tunneling rate at zero temperature and the thermal activation rate at finite temperature.}
    \label{fig:eff_potential}
\end{figure}

Substituting our field parameterization into the full field theory Hamiltonian derived from Eq.~(\ref{eq:totalenergy}) and performing the spatial integration, we obtain an effective one-dimensional Hamiltonian for the collective coordinate $\kappa(t)$:
\begin{equation}
H_{\rm{eff}} \; = \; \frac{m_{\rm {eff}}}{2}\dot{\kappa}^2 + V(\kappa) \, ,
\end{equation}
where the effective potential has the form:
\begin{equation}
\label{eq:effpot1}
V(\kappa) = E_{\rm{sph}} - \frac{1}{2}m_{\rm{eff}}\omega_{-}^2\kappa^2 + \frac{1}{4}\lambda_{\rm{eff}}m_{\rm{eff}}\kappa^4 + \mathcal{O}(\kappa^6) .
\end{equation}
The parameters in this effective model are determined from our stability analysis. The effective mass parameter $m_{\rm{eff}}$ is given by
\begin{equation}
    m_{\rm eff} \; = \; \frac{8 \pi}{g^2} \int_0^{\infty} r^2 dr \left(\eta_B^2 + \eta_C^2 + \eta_H^2 \right) \; = \; \frac{8 \pi}{g^2 m_{W}} \, ,
\end{equation}
and
\begin{equation}
\begin{aligned}
&\omega_{-}^2 \; = \; -m_W \int_0^{\infty} dr \left(1 - 2f(r)\right)^2 \left[ \eta_B(r)^2 + 2 \eta_C(r)^2 \right] \\
&+ m_W^2 r^2 h(r)^2 \left[ \eta_B(r)^2 + \eta_C(r)^2 \right] \\
&+ \left[ 2 f(r)^2 + \tfrac{1}{2} m_H^2 r^2 \left( h(r)^2 - 1 \right) \right] \eta_H(r)^2  \\
&+ 2\sqrt{2} (2f(r) - 1)\, \eta_B(r) \eta_C(r) 
- 2\sqrt{2} m_W r h(r)\, \eta_B(r) \eta_H(r)  \\
&- 2 m_W r^2 \eta_C(r) \eta_H(r) h'(r)
+ 2 m_W r^2 h(r) \eta_C(r) \eta_H'(r)  \\
&+ r^2 \left[ \eta_B'(r)^2 + \eta_H'(r)^2 \right] 
+ 2r\, \eta_B(r) \eta_B'(r) \\
&+ 2\sqrt{2} r (2f(r) - 1)\, \eta_C(r) \eta_B'(r) 
- 4\sqrt{2} r f'(r)\, \eta_B(r) \eta_C(r) \, ,
\end{aligned}
\end{equation}
and the effective quartic coupling $\lambda_{\rm eff}$ comes from the anharmonic terms in the energy functional:
\begin{equation}
\begin{aligned}
&\lambda_{\rm eff} \; = \; 2 m_W \int r^2  dr \left[
\tfrac{1}{2} \eta_B^2 \left( \eta_B^2 + \eta_H^2 + 4 \eta_C^2 \right) \right. \\
&\left.+ \tfrac{1}{2} \eta_C^2 \eta_H^2
+ \tfrac{m_H^2}{8 m_W^2} \eta_H^4
\right] \, .
\end{aligned}
\end{equation}
Our calculations yield $\lambda_{\rm eff} \simeq 3.1m_W^2$ and $\omega_{-}^2 \simeq -2.7 m_W^2$, consistent with the direct eigenvalue analysis presented in Section~\ref{sec:eigenvalueequations}. These results remain robust when using the analytical approximations for the unstable mode eigenfunctions derived in Section~\ref{sec:eigenvalueequations}, with deviations of less than 5\% compared to the exact numerical solutions. Combining these parameters with the sphaleron energy $E_{\rm sph} \simeq 3.8m_W/\alpha_W$ determined in Section~\ref{sec:staticminimsphandvac}, we can fully characterize the effective one-dimensional potential given by Eq.~\eqref{eq:effpot1}. This double-well potential, which provides a simplified representation of the energy landscape in configuration space along the unstable direction, has minima located at:
\begin{equation}
    \kappa_{\pm} \; = \; \pm \frac{\sqrt{-\omega_{-}^2}}{\sqrt{\lambda_{\rm eff}}} \; \simeq \; \pm 0.9 \, ,
\end{equation}
corresponding to the adjacent topological vacua on either side of the sphaleron barrier. Figure~\ref{fig:eff_potential} shows the effective potential as a function of the collective coordinate $\kappa$. The double-well structure clearly illustrates the energy barrier separating topologically distinct vacua, with the barrier height being precisely the sphaleron energy $E_{\rm{sph}} \simeq 9.1$ TeV at $\kappa = 0$.

\begin{figure}[t!]
    \centering

    % --- Top: kappa profile ---
    \begin{subfigure}[t]{0.45\textwidth}
        \centering
        \includegraphics[width=\textwidth]{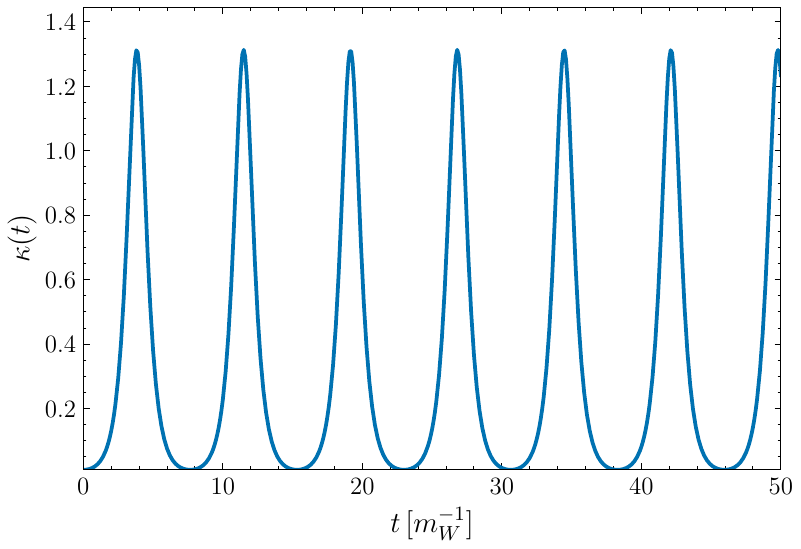}
    \end{subfigure}
    
    \vspace{1em}  % add vertical space

    % --- Middle: potential energy ---
    \begin{subfigure}[t]{0.45\textwidth}
        \centering
        \includegraphics[width=\textwidth]{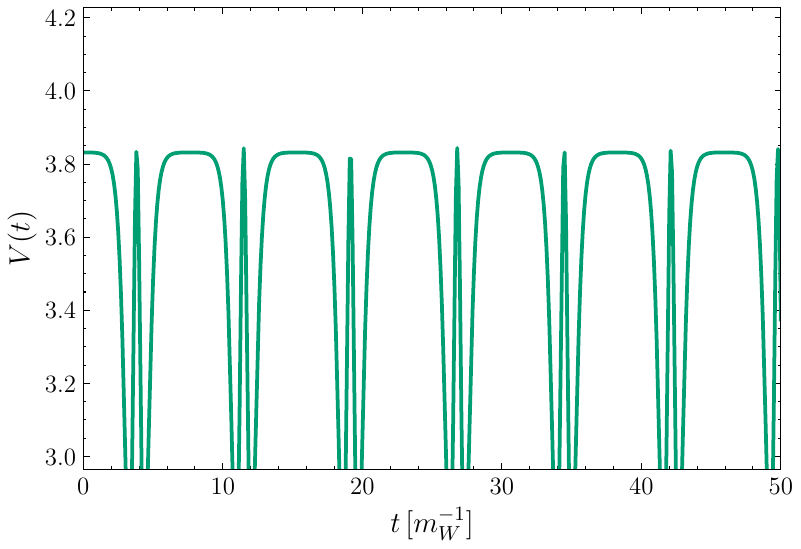}
    \end{subfigure}

    \vspace{1em}  % add vertical space

    % --- Bottom: kinetic energy ---
    \begin{subfigure}[t]{0.45\textwidth}
        \centering
        \includegraphics[width=\textwidth]{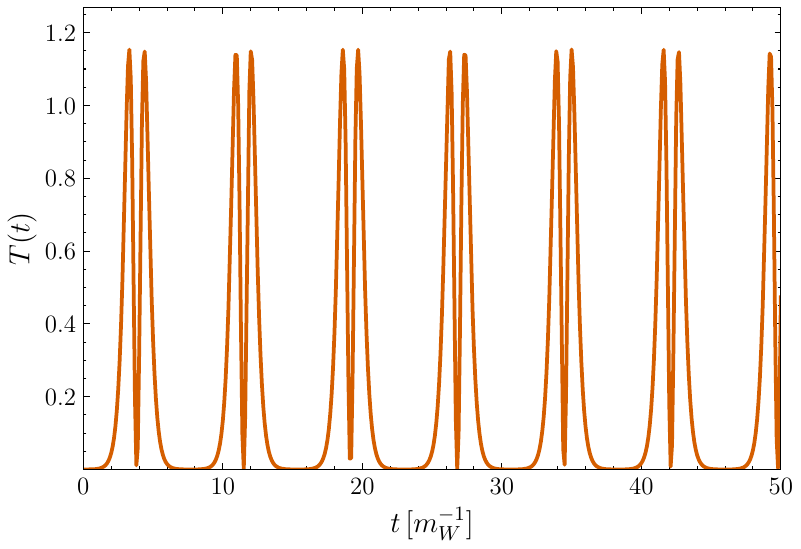}
    \end{subfigure}

    \caption{Time evolution of the sphaleron decay process: (Top) Collective coordinate $\kappa(t)$ showing initial exponential growth followed by oscillations around the vacuum state at $\kappa \simeq 0.9$. (Middle) Potential energy during the transition, displaying the system descent from the sphaleron barrier. (Bottom) Kinetic energy, showing the conversion between potential and kinetic energy as the system oscillates around the vacuum state. The decay is initiated with a small perturbation $\kappa(0) = 0.01$ and zero initial velocity. Time is measured in units of $m_W^{-1}$.}
    \label{fig:sphaleron_dynamics}
\end{figure}

The equation of motion for the collective coordinate $\kappa(t)$ follows from the effective Hamiltonian:
\begin{equation}
\kappa''(t) = 2.7\, m_W^2\, \kappa(t) - 3.1\, m_W^2\, \kappa(t)^3 \,.
\end{equation}
This nonlinear differential equation describes how the system evolves away from the sphaleron configuration. For small initial perturbations with $|\kappa| \ll 1$, the early-time evolution is dominated by the unstable mode's exponential growth:
\begin{equation}
\kappa(t) \simeq \kappa(0)\, e^{\sqrt{|\omega_{-}^2|}\, t} \simeq \kappa(0)\, e^{1.64\, m_W\, t} \,.
\end{equation}
As $\kappa$ grows, the nonlinear terms become important, eventually leading to oscillations around one of the potential minima. The final state depends on the sign of the initial perturbation $\kappa(0)$, determining which topological vacuum the system approaches.

Fig.~\ref{fig:sphaleron_dynamics} illustrates the time evolution of a sphaleron decay process initiated with a small perturbation $\kappa(0) = 0.01$ and zero initial velocity. The top panel shows the collective coordinate $\kappa(t)$ exhibiting initial exponential growth followed by oscillations around the minimum at $\kappa \simeq 0.9$. The middle and bottom panels display the corresponding evolution of potential and kinetic energy during the transition.

While this one-dimensional effective model captures the essential dynamics of sphaleron decay, it neglects interactions with other degrees of freedom that would lead to dissipation and eventual thermalization. In our companion paper \cite{MatchevVerner2025b}, we present full lattice simulations that account for these additional effects and provide a more comprehensive picture.
%%%%%%%%%%%%%%%%%%%%%%%%%%%%%%%%%%%%%%%%%%%%%%%%%%%%%%%%%%%%%%%%%%%%%
\section{Conclusions}
\label{sec:conclusions}
%%%%%%%%%%%%%%%%%%%%%%%%%%%%%%%%%%%%%%%%%%%%%%%%%%%%%%%%%%%%%%%%%%%%%
In this paper, we have presented a comprehensive analysis of sphaleron solutions in the electroweak theory, with emphasis on their stability properties and dynamics. Our investigation has systematically addressed several key aspects of sphaleron physics, providing both analytical insights and numerical precision needed for phenomenological applications.
We began by constructing static sphaleron solutions using spherically symmetric ansatz and numerical minimization techniques. Our calculations yielded the sphaleron energy $E_{\rm sph} \simeq 9.08 \, \rm{TeV}$, which sets the energy scale for baryon number violating processes. We then analyzed the minimum-energy path connecting topologically distinct vacua through the sphaleron, employing the constrained minimization approach of Akiba, Kikuchi, and Yanagida~\cite{Akiba:1988ay}. 

A central contribution of this work is our detailed stability analysis of the sphaleron configuration. By decomposing fluctuations into decoupled sectors and properly addressing gauge fixing issues, we identified precisely one unstable mode with eigenvalue $\omega_{-}^2 \simeq -2.7 \, m_W^2$. This negative eigenvalue confirms the saddle-point nature of the sphaleron and defines the unique direction in configuration space along which baryon number violating transitions proceed. Our analytical fits to the unstable mode profiles provide valuable tools for further theoretical and numerical investigations.

Building on the stability analysis, we developed an effective model for sphaleron dynamics using a collective coordinate approach. This reduction to a one-dimensional quantum mechanical system with a double-well potential captures the essential physics of sphaleron-mediated transitions while remaining computationally tractable. Our model characterizes the dynamics of sphaleron decay, including the initial exponential growth of perturbations and subsequent oscillations around topological vacua.

The results presented here have important implications for baryon number violation in both cosmological and high-energy contexts. In early universe cosmology, our detailed characterization of the sphaleron energy barrier and unstable mode improves predictions for thermal transition rates during the electroweak phase transition, with potential consequences for theories of baryogenesis. For high-energy physics, our analysis provides a foundation for evaluating potential sphaleron production cross-sections at future colliders and identifying distinctive experimental signatures.

Several avenues for future work emerge naturally from this investigation. First, extending the stability analysis to include finite-temperature effects would provide insights into how thermal fluctuations modify the sphaleron solution and energy barrier, particularly near the electroweak phase transition. Second, incorporating gauge and Higgs boson interactions with fermions would allow for explicit calculations of baryon and lepton number violating amplitudes in specific processes. Finally, generalizing the effective model to include dissipation and thermalization effects would enable more realistic simulations of sphaleron dynamics in complex environments.

In our companion paper \cite{MatchevVerner2025b}, we utilize the results presented here to perform real-time lattice simulations of sphaleron decay, directly studying the non-perturbative dynamics of baryon number violation. Together, these analyses advance our understanding of topology-changing processes in gauge theories and their phenomenological implications, bridging the gap between theoretical principles and potentially observable consequences of non-perturbative electroweak physics.

\begin{acknowledgments}
We thank Pierre Ramond for useful discussions. This work was supported in part by the U.S. Department of Energy award number DE-SC0022148. The work of KTM is supported in part by the Shelby Endowment for Distinguished Faculty at the University of Alabama. The work of S.V. was supported in part by DOE grant grant DE-SC0022148 at the University of Florida.
\end{acknowledgments}

\bibliography{references}

\end{document}